\documentclass[12pt]{article}

\usepackage{amsmath,amssymb,epsfig,bbold,color,subfigure,cite}

\normalbaselines

\begin{document}

\title{Complex oscillator and Painlev\'e IV equation}

\author{David J. Fern\'andez C. and J.C. Gonz\'alez \\ [5pt] Departamento de F\'isica, Cinvestav \\ A.P. 14-740, 07000 M\'exico D.F., Mexico}

\date{}

\maketitle

\begin{abstract}
Supersymmetric quantum mechanics is a powerful tool for generating exactly
solvable potentials departing from a given initial one. In this article the first- and second- order supersymmetric transformations will be used to obtain new exactly solvable potentials departing from the complex oscillator. The corresponding Hamiltonians turn out to be ruled by polynomial Heisenberg algebras. By applying a mechanism to reduce to second the order of these algebras, the connection with the Painlev\'{e} IV equation is achieved, thus giving place to new solutions for the Painlev\'{e} IV equation.
\end{abstract}

\bigskip


\bigskip


\section{Introduction}\label{**Sec-1}
As is well known, supersymmetric quantum mechanics (SUSY QM) is one of simplest techniques for generating new exactly solvable potentials departing from a given initial one \cite{AC04,Bag01,BS04,CKS01,Don07,Fer10,Fer05,GMR11,MR04,Suk05}. In general, the spectrum of the built Hamiltonian differs from the initial one in a finite number of levels. In the past this procedure has been applied mainly to hermitian Hamiltonians, in order to generate either real or complex new potentials \cite{AIC99,FMR03,RM03,Ros03,FG07,FGR07a,FR08}, with an associated spectrum which in the last case could contain a finite number of complex energy levels additional to the initial eigenvalues. The method has been used also for non-hermitian Hamiltonians with the so-called PT-symmetry \cite{BB98,Zno99,Mos02}. Indeed, in the context of optics it is possible to find families of complex potentials with entirely real spectra with a missing single level, the one associated to an arbitrary initial mode \cite{MMD13}. However, up to our knowledge there are not many works employing Hamiltonians with purely complex eigenvalues.
\\

Our aim in this work is to address this subject. In order to do that, we will take as initial system the so-called complex oscillator \cite{Bar86,JS86,BJ92,BJMP93}, which has the form of the harmonic oscillator potential but with a frequency $ \omega $ which is allowed now to be complex. We are going to show that some particularly interesting first-order SUSY transformations, which in the real case would introduce singularities in the generated potential, here will be non-singular and they will generate gaps in the equidistant spectrum of the complex oscillator.
\\

Let us recall that the SUSY partners of the harmonic oscillator potential supply explicit realizations of the even-order polynomial Heisenberg algebras (PHA) \cite{Fer84,FHN94,FNR95,FNO96,FH99,ACIN00,FNN04,CFNN04}. Moreover, there are subsets of this general family of SUSY partner potentials which are ruled simultaneously by second-order PHA. As the last are directly linked with the Painlev\'{e} IV (PIV) equation, then a simple technique is available for generating solutions to this equation departing from the extremal states of the Hamiltonian \cite{ACIN00,FNN04,CFNN04,MN08,Ma09,Ma10,BF11a,BF11b,Ber12,BF14}. In fact, plenty of real as well as complex solutions associated with real PIV parameters have been generated \cite{FNN04,CFNN04,MN08,Ma09,Ma10,BF11a,BF11b}. In addition, several complex solutions for complex PIV parameters have been built up \cite{Ber12}, all of them departing from the harmonic oscillator potential, which is real (for a review see, e.g. \cite{BF14}). However, since the complex oscillator can be mapped into the harmonic oscillator, thus we can generate complex solutions to the PIV equation directly by applying first-order supersymmetric transformations to the complex oscillator, and some special higher-order ones.
\\

This paper is organized as follows. In section \ref{**Sec-2} we will sketch the SUSY QM. Then, in section \ref{**Sec-3} we will discuss the PHA, with special emphasis placed in those of second-order, which have a direct link with the PIV equation. In section \ref{**Sec-4} we will generate the SUSY partners of the complex oscillator, while in section \ref{**Sec-5} we will study the corresponding solutions to the PIV equation. The final section contains our conclusions and some discussion concerning the perspectives of our work.

\section{Supersymmetric quantum mechanics}\label{**Sec-2}
Let us start with two Schr\"{o}dinger type Hamiltonians
	\begin{equation}\label{1}
H_{j}=-\frac{1}{2}\frac{\mathrm{d}^{2}}{\mathrm{d}x^{2}}+V_{j}\left(x\right)	,	\quad	j=0,1,
	\end{equation}
and suppose the existence of two first-order differential operators $ A_{1}^{\pm} $
	\begin{equation}\label{2}
A_{1}^{\pm}=\frac{1}{\sqrt{2}}\left[\mp\frac{\mathrm{d}}{\mathrm{d}x}+\beta_{1}\left(x,\varepsilon\right)\right]	,
	\end{equation}
which intertwine $ H_{0} $ with the new Hamiltonian $ H_{1} $ in the way \cite{Fer10,Fer05,Mie84,Suk85}
	\begin{equation}\label{3}
H_{1}A_{1}^{+}=A_{1}^{+}H_{0}	,	\quad	H_{0}A_{1}^{-}=A_{1}^{-}H_{1}	,
	\end{equation}
where $ \beta_{1}\left(x,\varepsilon\right) $ is an unknown complex function, so that $ A_{1}^{+} $ and $ A_{1}^{-} $ are not adjoint to each other.
\\

From Eqs.~(\ref{1})-(\ref{3}), the following relations have to be satisfied
	\begin{eqnarray}
& \beta_{1}^{\prime}\left(x,\varepsilon\right)+\beta_{1}^{2}\left(x,\varepsilon\right)	=	2\left[V_{0}\left(x\right)-\varepsilon\right]	,	\label{4}	\\
& V_{1}\left(x\right)	=	V_{0}\left(x\right)-\beta_{1}^{\prime}\left(x,\varepsilon\right) \label{5},
	\end{eqnarray}
with the same terminology as in the real case being used \cite{Fer84,FGN98,Ros98}, i.e., $ \varepsilon\in\mathbb{C} $ is the so-called factorization energy. Through the substitution
	\begin{equation}\label{6}
\beta_{1}\left(x,\varepsilon\right)=\left[\ln{u\left(x,\varepsilon\right)}\right]^{\prime}=\frac{{u}^{\prime}\left(x,\varepsilon\right)}{u\left(x,\varepsilon\right)}	,
	\end{equation}
Eq.~(\ref{4}) is transformed into
	\begin{equation}\label{7}
H_{0}\, u\left(x,\varepsilon\right)=\varepsilon \, u\left(x,\varepsilon\right)	,
	\end{equation}
namely, $ u\left(x,\varepsilon\right) $ is a solution of the stationary Schr\"{o}dinger equation for $ H_{0} $ associated with $ \varepsilon $.
\\

It is interesting to observe that the involved Hamiltonians become factorized as
	\begin{equation}\label{8}
H_{0}=A_{1}^{-}A_{1}^{+}+\varepsilon	,	\quad	H_{1}=A_{1}^{+}A_{1}^{-}+\varepsilon	.
	\end{equation}
From the intertwining relationships (\ref{3}), we can obtain the eigenfunctions $ \psi^{\left(1\right)}_{n} $ of $ H_{1} $ and their corresponding eigenvalues, departing from the normalized eigenfunctions $ \psi^{\left(0\right)}_{n} $ of $ H_{0} $ with eigenvalues $ E_{n} $:
	\begin{subequations}
	\begin{align}
\psi^{\left(1\right)}_{n}\left(x\right)	&	=	C_{n}A_{1}^{+}\psi^{\left(0\right)}_{n}\left(x\right)	,	\label{9a}	\\
H_{1}\psi^{\left(1\right)}_{n}\left(x\right)	&	=	E_{n}\psi^{\left(1\right)}_{n}\left(x\right)	,	\quad	n=0,1,2,\cdots ,	\label{9b}
	\end{align}
	\end{subequations}
which means that $ \psi^{\left(0\right)}_{n} $ and $ \psi^{\left(1\right)}_{n} $ share the eigenvalues $ E_{n} $. The normalization factors $ C_{n} $ in Eq.~(\ref{9a}) cannot be straightforwardly obtained, as for real $ u $, since now $ A_{1}^{+}\neq (A_{1}^{-})^{\dagger} $.
\\

Note that there exists a wavefunction annihilated by $ A_{1}^{-} $ given by
	\begin{equation}\label{10}
\psi_{\varepsilon}^{\left(1\right)}\left(x\right)\propto\frac{1}{u\left(x,\varepsilon\right)}	,
	\end{equation}
which satisfies $ H_{1} \, \psi_{\varepsilon}^{\left(1\right)}=\varepsilon\, \psi_{\varepsilon}^{\left(1\right)} $. If $ \psi_{\varepsilon}^{\left(1\right)} $ can be normalized, it becomes also an eigenfunction of $ H_{1} $ with eigenvalue $ \varepsilon $; in such a case, the spectrum of $ H_{1} $ is the same as for $ H_{0} $ plus an extra energy level at $ \varepsilon $:
	\begin{equation*}
\mathsf{Sp}\left(H_{1}\right)=\mathsf{Sp}\left(H_{0}\right)\cup\lbrace\varepsilon\rbrace	.		\end{equation*}

In order to avoid singularities in the new potential $ V_{1} $, the transformation function $ u\left(x,\varepsilon\right) $ (also called seed solution), must not have zeros. This is simple to achieve due to the complex nature of the function itself: hardly the real and imaginary parts of $ u\left(x,\varepsilon\right) $ will vanish simultaneously at the same point. In addition, in this case the oscillation theorem does not hold anymore \cite{BS91}. Due to this, from now on we are going to suppose that $ u\left(x,\varepsilon\right) $ has not zeros on the real axis.
\\

By iterating now $ k $ times the first-order transformation just discussed, a $ k $-th order complex SUSY transformation is generated, where $ k $ new energy levels can be created so that the spectrum of the corresponding Hamiltonian becomes \cite{Fer05,FGN98,Ros98,AIS93,MNR00}
	\begin{equation}\label{11}
\mathsf{Sp}\left(H_{k}\right)=\lbrace\varepsilon_{1},\cdots ,\varepsilon_{k},E_{n};\;n=0,1,\cdots\rbrace	.
	\end{equation}
The standard SUSY algebra with two generators $ Q_{1} $, $ Q_{2} $ introduced by Witten \cite{Wit81} (see also \cite{Nie84})
	\begin{eqnarray}
& \left[Q_{j},H_{ss}\right]	=	Q_{j}H_{ss}-H_{ss}Q_{j}=0	,	\nonumber	\\
& \lbrace Q_{j},Q_{k}\rbrace =	Q_{j}Q_{k}+Q_{k}Q_{j}=\delta_{jk}H_{ss}	,	\quad	j,k=1,2,	\label{12}
	\end{eqnarray}
can now be realized in the way
	\begin{equation*}
Q_{1}=\frac{1}{\sqrt{2}}\left(Q^{+}+Q^{-}\right)	,	\quad	Q_{2}=\frac{1}{i\sqrt{2}}\left(Q^{+}-Q^{-}\right)	,
	\end{equation*}
where
	\begin{equation*}
Q^{+}=\begin{pmatrix}
0	&	B_{k}^{+}	\\	0	&	0	\end{pmatrix}	,	\quad	Q^{-}=\begin{pmatrix}
0	&	0	\\	B_{k}^{-}	&	0	\end{pmatrix}	,
	\end{equation*}
	\begin{equation}\label{13}
H_{ss}=\lbrace Q^{-},Q^{+}\rbrace =\begin{pmatrix}
B_{k}^{+}B_{k}^{-}		&	0	\\
0						&	B_{k}^{-}B_{k}^{+}
\end{pmatrix}
	=	\left(H_{\mathrm{d}}-\varepsilon_{1}\right)\cdots\left(H_{\mathrm{d}}-\varepsilon_{k}\right)	,	
	\end{equation}
	\begin{equation*}
H_{\mathrm{d}}=\begin{pmatrix}
H_{k}	&	0	\\	0	&	H_{0}	\end{pmatrix}	.
	\end{equation*}
The so-called supersymmetric Hamiltonian $ H_{ss} $ is a polynomial of degree $ k $-th in $ H_{\mathrm{d}} $, since $ H_{0} $ and $ H_{k} $ are intertwined as
	\begin{equation}\label{14}
H_{k}B_{k}^{+}=B_{k}^{+}H_{0}	,	\quad	H_{0}B_{k}^{-}=B_{k}^{-}H_{k}	,
	\end{equation}
with
	\begin{equation}\label{15}
B_{k}^{+}=A_{k}^{+}\cdots A_{1}^{+}	,	\quad	B_{k}^{-}=A_{1}^{-}\cdots A_{k}^{-}	,
	\end{equation}
being $ k $-th order differential intertwining operators. The initial and final potentials, $ V_{0} $ and $ V_{k} $ respectively, are related by
	\begin{align}
V_{k}\left(x\right)	&	=	V_{0}\left(x\right)-\left[\ln{W\left(u_{1},\cdots ,u_{k}\right)}\right]^{\prime\prime}	,	\label{16}	\\
u_{j}	&	=u_{j}\left(x\right)=:	u\left(x,\varepsilon_{j}\right)	,	\quad	j=1,\cdots ,k	\label{17}	,
	\end{align}
where $ W\left(u_{1},\cdots ,u_{k}\right) $ is the Wronskian of the $ k $ Schr\"{o}dinger seed solutions $ \left\lbrace u_{1},\cdots ,u_{k}\right\rbrace $. The eigenfunctions of $H_k$ associated to the eigenvalues $E_n$ are given by:
	\begin{equation}\label{18}
\psi^{\left(k\right)}_{n}\left(x\right)=C_{n}B_{k}^{+}\psi^{\left(0\right)}_{n}\left(x\right)	,	\quad n=0,1,2,\cdots	
	\end{equation}
On the other hand, those associated to the new eigenvalues $\varepsilon_{j}$ become:
	\begin{equation}\label{19}
\psi^{\left(k\right)}_{\varepsilon_{j}}\left(x\right)\propto\frac{\widetilde W_k\left(u_{j}\right)}{W\left(u_{1},\cdots ,u_{k}\right)}	,	\quad j=1,\cdots ,k,
	\end{equation}
where $\widetilde W_k\left(u_{j}\right)$ denotes the Wronskian of the seed solutions $\left\lbrace u_{1},\cdots ,u_{k}\right\rbrace$ excluding
the $j$th one ($u_{j}$). In particular, for $k=1$ the first order case discussed previously is recovered if we assume that $ W\left(u_{1}\right)\equiv u_{1}$ and $ \widetilde{W}_1\left(u_{1}\right)\equiv 1$ (compare Eqs.~(\ref{5}),(\ref{9a}),(\ref{10}) with Eqs.~(\ref{16}),(\ref{18}),(\ref{19})).
\\

Let us note that this technique has been employed successfully to generate new solvable potentials $ V_{k} $ departing from a given initial one $ V_{0} $ for several interesting physical Hamiltonians (mainly Hermitian \cite{Bag01,CKS01,Fer10,MR04,Fer84,FGN98,Ros98}).

\section{Polynomial Heisenberg algebras}\label{**Sec-3}

A $ \left(m-1\right) $-th order PHA is a deformation of the Heisenberg-Weyl algebra with three generators $ \lbrace H,\mathcal{L}_{m}^{+},\mathcal{L}_{m}^{-}\rbrace $ which obey two standard commutation relations
	\begin{subequations}
	\begin{equation}\label{20a}
\left[H,\mathcal{L}_{m}^{\pm}\right]=\pm\mathcal{L}_{m}^{\pm}	,
	\end{equation}
and an atypical commutator characterizing the deformation:
	\begin{equation}\label{20b}
\left[\mathcal{L}_{m}^{-},\mathcal{L}_{m}^{+}\right]=N_{m}\left(H+1\right)-N_{m}\left(H\right)=P_{m-1}\left(H\right)	,
	\end{equation}
	\end{subequations}
where $ \mathcal{L}_{m}^{\pm} $ are $ m $-th order differential ladder operators, $ N_{m}\left(H\right)=\mathcal{L}_{m}^{+}\mathcal{L}_{m}^{-} $ is a polynomial in $ H $ of degree $ m $ representing a generalization of the number operator, so that $ P_{m-1}\left(H\right) $ is a polynomial in $ H $ of degree $ m-1 $. The generalized number operator can be factorized as
	\begin{equation}\label{21}
N_{m}\left(H\right)=\prod_{j=1}^{m}{\left(H-\mathcal{E}_{j}\right)}	,
	\end{equation}
$ \mathcal{E}_{j} $ being the zeros of the polynomial $ N_{m}\left(H\right) $, which correspond to the associated extremal state energies.
\\

\begin{figure}[!h]
	\centering
	\begin{tabular}{cc}
	\subfigure[]{\includegraphics[width=0.43\textwidth]{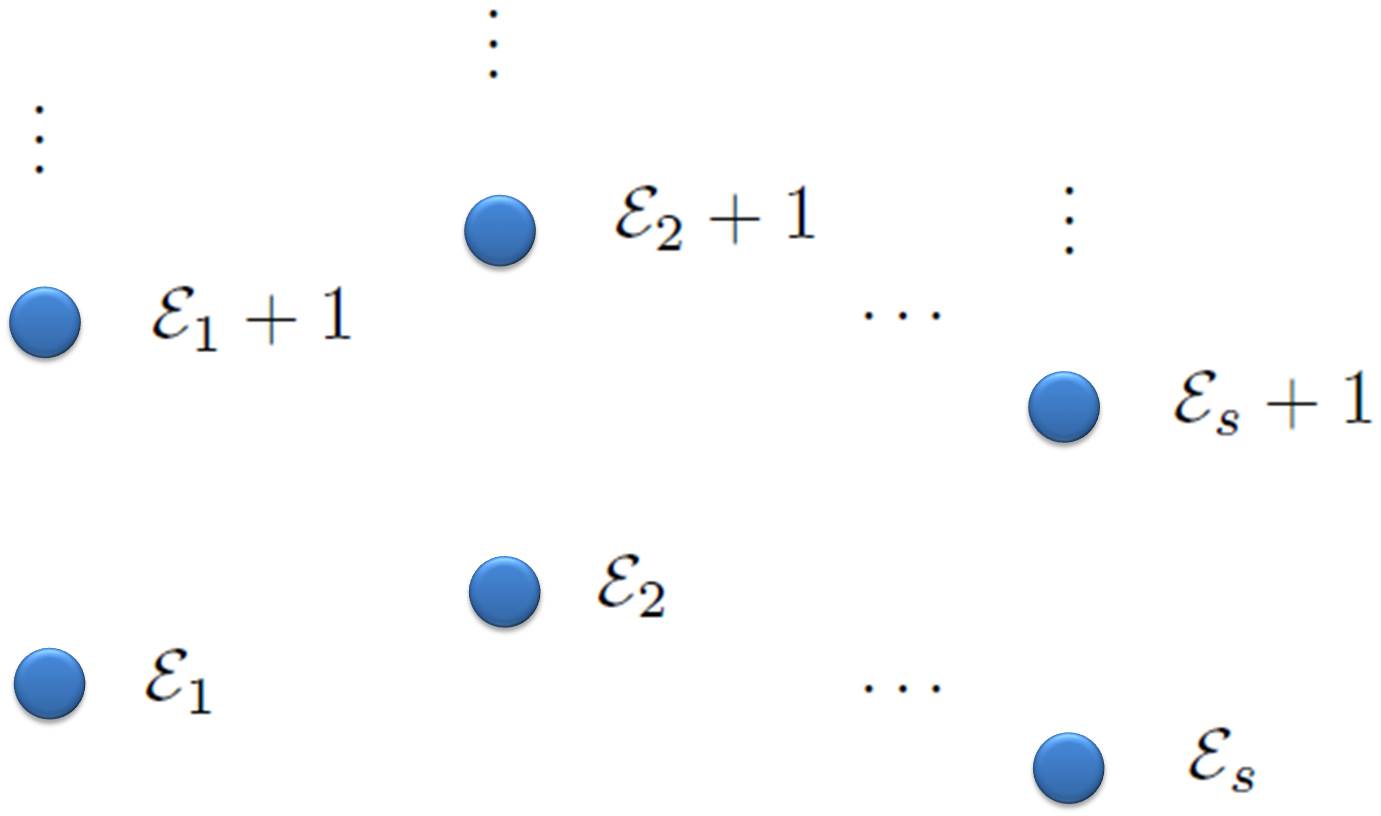}}\hspace{1.4cm}
	&	\subfigure[]{\includegraphics[width=0.43\textwidth]{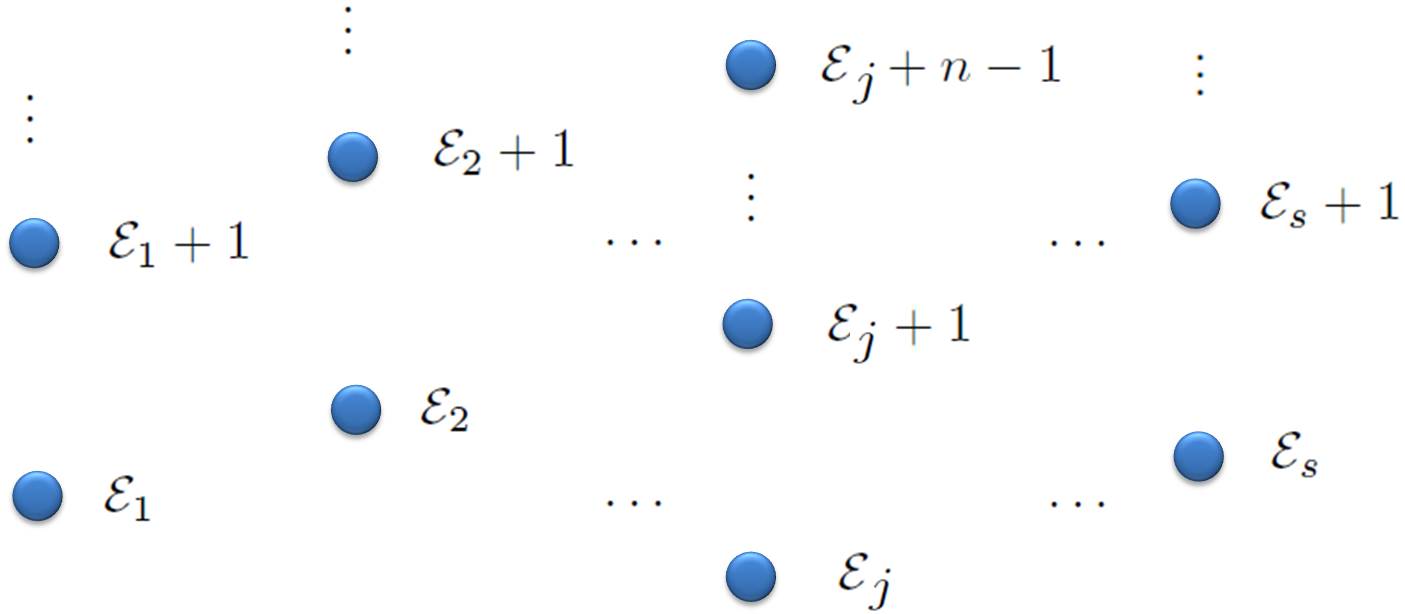}}		\\
	\end{tabular}
		\caption{Spectrum of the Hamiltonian $ H $ having either: (a) $ s $ physical extremal states from which $ s $ independent endless ladders are generated, or (b) $ s $ physical extremal states, $ s-1 $ of which are associated to endless ladders and a finite one with $ n $ steps.}
		\label{--fig:1}
\end{figure}

Let us consider the kernel $ K^{-} $ of $ \mathcal{L}_{m}^{-} $, i.e., the $ m $-dimensional space of solutions of the $ m $-th order differential equation
	\begin{equation}\label{22}
\mathcal{L}_{m}^{-}\psi =0	,
	\end{equation}
which is invariant under $ H $. Then, from Eqs.~(\ref{21}) and (\ref{22}) we obtain
	\begin{equation}\label{23}
\mathcal{L}_{m}^{+}\mathcal{L}_{m}^{-}\psi =\prod_{j=1}^{m}{\left(H-\mathcal{E}_{j}\right)}\psi =0	.
	\end{equation}
We choose as a natural basis of $ K^{-} $ the solutions of the stationary Schr\"{o}dinger equation for $ H $ associated to $ \mathcal{E}_{j}\in\mathbb{C} $:
	\begin{equation}\label{24}
H\psi_{\mathcal{E}_{j}}=\mathcal{E}_{j}\psi_{\mathcal{E}_{j}}	.
	\end{equation}
Departing from these so-called extremal states, by acting iteratively with $ \mathcal{L}_{m}^{+} $ onto them, we can build $ m $ mathematical energy ladders with the same spacing $ \Delta E=1 $ starting from each $ \mathcal{E}_{j} $. If $ s $ of these states are normalizable, $ \lbrace\psi_{\mathcal{E}_{j}};\: j=1,\cdots ,s\rbrace $, then $ \mathsf{Sp}\left(H\right) $ will have $ s $ independent endless physical ladders (see Fig. \ref{--fig:1}a). It could happen, however, that for some $ j\in\left\lbrace 1,\cdots ,s\right\rbrace $
	\begin{equation}\label{25}
\left(\mathcal{L}_{m}^{+}\right)^{n-1}\psi_{\mathcal{E}_{j}}\neq 0	,	\quad	\left(\mathcal{L}_{m}^{+}\right)^{n}\psi_{\mathcal{E}_{j}}=0	,	\quad	n\in\mathbb{N}	.
	\end{equation}
Then, by analyzing the expression $ \mathcal{L}_{m}^{-}(\mathcal{L}_{m}^{+})^{n}\psi_{\mathcal{E}_{j}}=0$, we observe that one of the remaining non-physical roots must satisfy $ \mathcal{E}_{l}=\mathcal{E}_{j}+n $, $ l\in \left\lbrace s+1,\cdots ,m\right\rbrace $. Hence, $ \mathsf{Sp}\left(H\right) $ will have $ s-1 $ endless ladders and a finite one of length $ n $, which starts from $ \mathcal{E}_{j} $ and ends at $ \mathcal{E}_{j}+n-1 $ (see Fig. \ref{--fig:1}b).

It is important to identify now the most general one-dimensional Schr\"{o}dinger Hamiltonians which are ruled by the polynomial Heisenberg algebras of Eqs.~(18) for growing values in the order $ m $ of the ladder operators $ \mathcal{L}_{m}^{\pm} $ \cite{CFNN04}. The answer runs as follows: for $ m=1 $ and $ m=2 $ (ladder operators of first- and second- orders respectively) the most general potentials have a closed form, becoming the harmonic and effective radial oscillators respectively. On the other hand, for $ m=3 $ (third-order ladder operators) there is not a closed expression for the potential, but it depends on a function $ g $ which must satisfy the PIV equation. Because its importance for this work, let us review quickly this last case.
	\begin{enumerate}
\item[3.1] Second order PHA.
\\
The connection between PIV equation and PHA appears for $ m=3 $ \cite{CFNN04,Nie84}. Thus, let $ \mathcal{L}_{3}^{+} $ be a third-order differential ladder operator factorized as the following product of a first- and a second-order differential one:
	\begin{subequations}
	\begin{align}
\mathcal{L}_{3}^{+}	&	=	\mathcal{L}_{1}^{+}\mathcal{L}_{2}^{+}	,	\label{26a}	\\
\mathcal{L}_{1}^{+}	&	=	\frac{1}{\sqrt{2}}\left[-\frac{\mathrm{d}}{\mathrm{d}x}+f\left(x\right)\right]	,	\label{26b}	\\
\mathcal{L}_{2}^{+}	&	=	\frac{1}{2}\left[\frac{\mathrm{d}^{2}}{\mathrm{d}x^{2}}+g\left(x\right)\frac{\mathrm{d}}{\mathrm{d}x}+h\left(x\right)\right]	,	\label{26c}
	\end{align}
	\end{subequations}
$ f $, $ g $ and $ h $ being complex functions to be determined. We assume the existence of an auxiliary Hamiltonian $ H_{\mathrm{a}} $ which is intertwined with $ H $ in the way
	\begin{equation}\label{27}
H\mathcal{L}_{1}^{+}=\mathcal{L}_{1}^{+}\left(H_{\mathrm{a}}+1\right)	,	\quad	H_{\mathrm{a}}\mathcal{L}_{2}^{+}=\mathcal{L}_{2}^{+}H	.
	\end{equation}
After some work, the resulting uncoupled system of equations becomes
	\begin{subequations}
	\begin{align}
f	&	=	x+g	,	\label{28a}	\\
h	&	=	-\frac{g^{\prime}}{2}+\frac{g^{2}}{2}-2V+\mathcal{E}_{1}+\mathcal{E}_{2}-2	,	\label{28b}	\\
V	&	=	\frac{x^{2}}{2}-\frac{g^{\prime}}{2}+\frac{g^{2}}{2}+xg+\mathcal{E}_{3}-\frac{1}{2}	,	\label{28c}
	\end{align}
	\end{subequations}
where $ g $ satisfies
	\begin{subequations}
	\begin{align}
g^{\prime\prime}	&	=	\frac{{g^{\prime}}^{2}}{2g}+\frac{3}{2}g^{3}+4xg^{2}+2\left(x^{2}-a\right)g+\frac{b}{g}	,	\label{29a}	\\
a	&	:=\mathcal{E}_{1}+\mathcal{E}_{2}-2\mathcal{E}_{3}-1	,	\quad	b:=-2\left(\mathcal{E}_{1}-\mathcal{E}_{2}\right)^{2}	,	\label{29b}
	\end{align}
	\end{subequations}
which is the PIV equation with parameters $ a,b\in\mathbb{C} $ \cite{IKSY91}. Note that it is essential to know the function $ g $ associated to a pair of parameters $ a $, $ b $; then, the extremal energies $ \mathcal{E}_{j} $, which now are complex, as well as the potential $ V $ in Eq.~(\ref{28c}) can be found. Moreover, the three extremal states become expressed in terms of $ g $ as:
	\begin{subequations}
	\begin{align}
\psi_{\mathcal{E}_{1}}	&	\propto	\left(\frac{g^{\prime}}{2g}-\frac{g}{2}-\frac{1}{g}\sqrt{-\frac{b}{2}}-x\right)e^{\int{\left(\frac{g^{\prime}}{2g}+\frac{g}{2}-\frac{1}{g}\sqrt{-\frac{b}{2}}\right)\;\mathrm{d}x}}	,	\label{30a}	\\
\psi_{\mathcal{E}_{2}}	&	\propto	\left(\frac{g^{\prime}}{2g}-\frac{g}{2}+\frac{1}{g}\sqrt{-\frac{b}{2}}-x\right)e^{\int{\left(\frac{g^{\prime}}{2g}+\frac{g}{2}+\frac{1}{g}\sqrt{-\frac{b}{2}}\right)\;\mathrm{d}x}}	,	\label{30b}	\\
\psi_{\mathcal{E}_{3}}	&	\propto	e^{-\frac{1}{2}x^{2}-\int{g\;\mathrm{d}x}}	.	\label{30c}
	\end{align}
	\end{subequations}

In addition, the generalized number operator is cubic in the Hamiltonian:
	\begin{equation}\label{31}
N_{3}\left(H\right)=\left(H-\mathcal{E}_{1}\right)\left(H-\mathcal{E}_{2}\right)\left(H-\mathcal{E}_{3}\right)	.
	\end{equation}

Conversely, if we are able to identify Hamiltonians having third-order differential ladder operators, it is possible to design a simple mechanism to obtain solutions to the PIV equation (see \cite{CFNN04} and section \ref{**Sec-5} in this paper).
	\end{enumerate}


\section{SUSY partners of the complex oscillator}\label{**Sec-4}

In order to implement the SUSY technique, we need to solve first the stationary Schr\"{o}dinger equation for the complex oscillator potential:
	\begin{equation}\label{32}
V_{0}\left(x\right)=\frac{1}{2}\omega^{2}x^{2}	,	\quad	\omega=e^{i\theta}	,	\quad	-\frac{\pi}{2}\leq\theta <\frac{3\pi}{2}	,
	\end{equation}
with an arbitrary complex factorization energy $ \varepsilon $, $ \omega $ being a dimensionless complex frequency and $ \theta $ its phase. A direct calculation leads to \cite{JR98}
	\begin{align}
u\left(x,\varepsilon\right)	&	=	e^{-\frac{1}{2}\omega x^{2}}\left[_{1}F_{1}\left(\frac{1}{4}-\frac{\varepsilon}{2\omega},\frac{1}{2};\omega x^{2}\right)+\lambda\;x\,_{1}F_{1}\left(\frac{3}{4}-\frac{\varepsilon}{2\omega},\frac{3}{2};\omega x^{2}\right)\right]	\nonumber	\\
	&	=	e^{\frac{1}{2}\omega x^{2}}\left[_{1}F_{1}\left(\frac{1}{4}+\frac{\varepsilon}{2\omega},\frac{1}{2};-\omega x^{2}\right)+\lambda\;x\,_{1}F_{1}\left(\frac{3}{4}+\frac{\varepsilon}{2\omega},\frac{3}{2};-\omega x^{2}\right)\right]	,	\label{33}
	\end{align}
where we will take
	\begin{equation*}
\lambda =2\nu\frac{\Gamma\left(\frac{3}{4}-\frac{\varepsilon}{2\omega}\right)}{\Gamma\left(\frac{1}{4}-\frac{\varepsilon}{2\omega}\right)}	,	\quad	\left|\nu\right|<1	,
	\end{equation*}
and $ _{1}F_{1} $ is the confluent hypergeometric function (Kummer). In general, $ u\left(x,\varepsilon\right) $ is a divergent solution of the Schr\"{o}dinger equation since it does not satisfy $ \lim_{\vert x\vert\rightarrow \infty}{\vert u\left(x,\varepsilon\right)\vert=0} $. To ensure its square-integrability, any of the two series $ _{1}F_{1} $ in (\ref{33}) must reduce to a polynomial. Thus, depending on the value of $ \theta $, three different cases are identified:
	\begin{enumerate}
	\item[\emph{(i)}] For $ -\frac{\pi}{2}<\theta <\frac{\pi}{2} $ the energy levels of the complex oscillator turn out to be given by
	\begin{equation}\label{34}
E_{n}\left(\theta\right)=\left(n+\frac{1}{2}\right)e^{i\theta}	,	\quad	n=0,1,2,\cdots	,
	\end{equation}
and the corresponding eigenfunctions read
	\begin{equation}\label{35}
\phi_{n}\left(x\right)=C_{n}\mathrm{H}_{n}(\sqrt{\omega}x)e^{-\frac{1}{2}\omega x^{2}}	,	\quad	\sqrt{\omega}=e^{i\frac{\theta}{2}}	,
	\end{equation}
where $ C_{n} $ are normalization factors and $ \mathrm{H}_{n}(\sqrt{\omega}x) $ are Hermite polynomials of the complex argument $\sqrt{\omega}x$. It is worth to note that the Hermite polynomials of complex argument appear in several physically interesting situations, e.g., in the calculation of the photon distribution for squeezed states and their Wigner distribution function \cite{MS11}, in the eigenfunctions analysis for an optical beam in the paraxial approximation \cite{Si73}, among others \cite{BF13}. Let us note also that the Hermite polynomials of complex argument, in general, are not longer orthogonal \cite{Si73}.

	\item[\emph{(ii)}] For $ \frac{\pi}{2}<\theta <\frac{3\pi}{2} $ it turns out that
	\begin{equation}\label{36}
E_{n}\left(\theta\right)=\left(n+\frac{1}{2}\right)e^{i\left(\theta -\pi\right)}	,	\quad	n=0,1,2,\cdots	,
	\end{equation}
while the corresponding eigenfunctions become
	\begin{equation}\label{37}
\phi_{n}\left(x\right)=D_{n}\mathrm{H}_{n}(\sqrt{-\omega}x)e^{\frac{1}{2}\omega x^{2}}	,	\quad	\sqrt{-\omega}=e^{i\frac{\theta -\pi}{2}}	.
	\end{equation}

	\item[\emph{(iii)}] For $ \theta =\pm\frac{\pi}{2} $ there are not square-integrable eigenfunctions of $ H $ since now the potential corresponds to the repulsive oscillator \cite{BF13}.
	\end{enumerate}
Thus, the eigenvalues of the complex oscillator potential lie always either in the first or in the fourth quadrant on the complex energy plane. This result is consistent with the fact that our potential is invariant under the symmetry transformation $ \omega\rightarrow -\omega $ and, therefore, the eigenfunctions and eigenvalues for $ \omega $ and $ -\omega $ have to be the same, which is indeed the case. As a consequence, we can restrict ourselves to the domain $ -\frac{\pi}{2}<\theta <\frac{\pi}{2} $. Moreover, since
	\begin{equation}\label{38}
E_{n}\left(-\theta\right)=\left[E_{n}\left(\theta\right)\right]^{\star}	,
	\end{equation}
it is enough to work with the domain $ \theta\in\left[0,\frac{\pi}{2}\right) $, which we will do from now on.
\\

In analogy to the harmonic oscillator, it is possible to introduce now annihilation and creation operators $ a_{\omega}^{\pm} $, which define a fixed direction $ \omega $ on the complex energy plane, in the following way:
	\begin{equation}\label{39}
a_{\omega}^{\pm}:=\frac{1}{\sqrt{2}}\left(\mp\frac{\mathrm{d}}{\mathrm{d}x}+\omega x\right)	.
	\end{equation}
Since $ \omega\in\mathbb{C} $, then $ a_{\omega}^{+} $ and $ a_{\omega}^{-} $ are not in general adjoint to each other.
The algebra of these operators is analogous to the harmonic oscillator one, i.e., the Heisenberg-Weyl's algebra \cite{BHP97}:
	\begin{subequations}
	\begin{align}
\left[a_{\omega}^{-},a_{\omega}^{+}\right]	&	=	\omega	,	\label{40a}	\\
\left\lbrace a_{\omega}^{-},a_{\omega}^{+}\right\rbrace	&	=	2H_{0}	,	\label{40b}	\\
\left[H_{0},a_{\omega}^{\pm}\right]	&	=	\pm\omega a_{\omega}^{\pm}	.	\label{40c}
	\end{align}
	\end{subequations}
Let us note that the eigenfunctions and eigenvalues for the complex oscillator can be found in an elegant alternative way by using the operators $ a_{\omega}^{\pm} $, as for the harmonic oscillator. The square-integrable eigenfunctions turn out to be
	\begin{align}
\phi_{n}\left(x\right)	&	=	\frac{\widetilde C_{n}}{\sqrt{n!}}\left(a_{\omega}^{+}\right)^{n}\phi_{0}\left(x\right)	\nonumber	\\
	&	=	C_{n}\mathrm{H}_{n}(\sqrt{\omega}x)e^{-\frac{1}{2}\omega x^{2}},	\label{41}
	\end{align}
where $ \phi_{0}\propto e^{-\frac{1}{2}\omega x^{2}} $ is the ground state. Note that when $ \varepsilon =E_{n}\left(\theta\right)=:E_{n} $, the general solution (\ref{33}) can be reduced to (\ref{41}), i.e., $ u\left(x,E_{n}\right)=\phi_{n}\left(x\right) $.
\\

\begin{figure}[!h]
	\centering
	\includegraphics[width=0.45\textwidth]{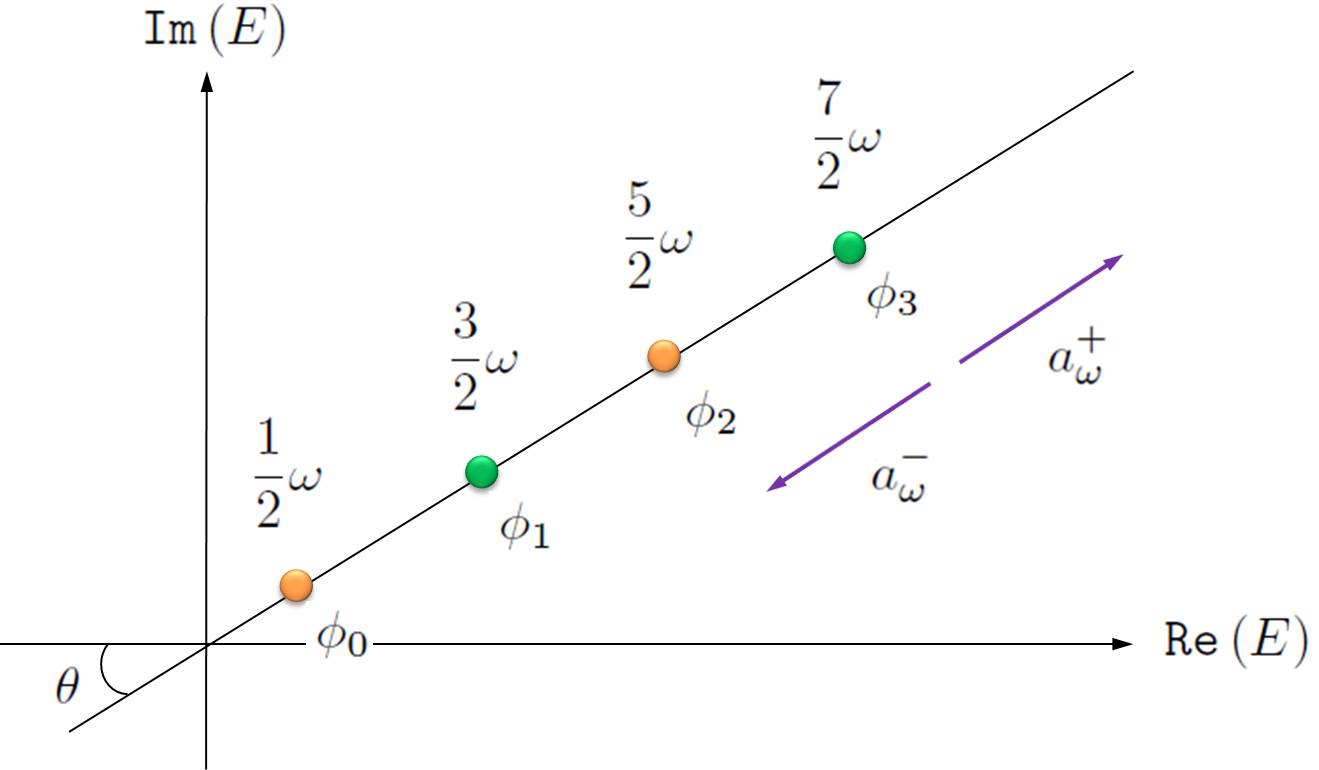}
		\caption{Diagram of the complex energy plane for a fixed $ \theta $. Some eigenvalues are shown at their right positions.}
		\label{--fig:2}
\end{figure}

A diagram of the complex energy plane, the eigenvalues $ E_{n} $ of the complex oscillator for a fixed $ \theta $ as well as the action of the annihilation and creation operators $ a_{\omega}^{\pm} $ can be seen in Fig. \ref{--fig:2}.

Let us apply now a non-singular $ k $-th order supersymmetric transformation which creates $ k $ new levels additional to $ E_{n} $ for the new Hamiltonian $ \mathsf{H}_{k}=-\frac{1}{2}\frac{\mathrm{d}^{2}}{\mathrm{d}x^{2}}+V_{k} $, i.e.
	\begin{equation*}
\mathsf{Sp}\left(\mathsf{H}_{k}\right)=\left\lbrace\varepsilon_{j},E_{n};\; j=1,\cdots ,k;\; n=0,1,2,\cdots\right\rbrace	.
	\end{equation*}
In order to do that, we need to take either $ k $ solutions $ \lbrace\beta_{1}\left(x,\varepsilon_{j}\right);\;j=1,\cdots ,k\rbrace $ of the initial Riccati equation for $ k $ different complex factorization energies $ \varepsilon_{j} $, or $ k $ solutions $ \lbrace u_{j};\;j=1,\cdots ,k\rbrace $ of the corresponding Schr\"{o}dinger equation. Since the oscillation theorem is not longer valid for solutions of the stationary Schr\"{o}dinger equation associated to complex potentials, then the factorization energies $ \left\lbrace\varepsilon_{1},\cdots ,\varepsilon_{k}\right\rbrace $ can be chosen essentially at any position on the complex energy plane. This fact will be quite useful later on, when we will explore some particularly interesting (and simple) SUSY transformations. The new potential $ V_{k} $ is given by Eq.~(\ref{16}) with $ V_{0}\left(x\right)=\frac{1}{2}\omega^{2}x^{2} $.
\\

The algebra for systems described by $ \mathsf{H}_{k} $ can be built departing from the complex oscillator one. Indeed, the natural ladder operators associated to $ \mathsf{H}_{k} $ are given by
	\begin{equation}\label{42}
\mathcal{L}_{2k+1}^{\pm}=B_{k}^{+}a_{\omega}^{\pm}B_{k}^{-}	,
	\end{equation}
where the $ k $-th order differential intertwining operators $ B_{k}^{\pm} $ are given in Eq.~(\ref{15}). It turns out that the standard commutation relations of the PHA are satisfied:
	\begin{equation}\label{43}
\left[\mathsf{H}_{k},\mathcal{L}_{2k+1}^{\pm}\right]=\pm\omega\mathcal{L}_{2k+1}^{\pm}	,
	\end{equation}
which differ from Eq.~(\ref{20a}) in the factor $ \omega $. In order to stick to the convention of Eq.~(\ref{20a}), let us redefine the Hamiltonian as
	\begin{equation}\label{44}
H_{k}=\frac{\mathsf{H}_{k}}{\omega}	,
	\end{equation}
so that the relationship (\ref{43}) is transformed into
	\begin{equation}\label{45}
\left[H_{k},\mathcal{L}_{2k+1}^{\pm}\right]=\pm\mathcal{L}_{2k+1}^{\pm}	.
	\end{equation}
The Hamiltonians $ H_{k} $ associated with this algebra take the form
	\begin{equation}\label{46}
H_{k}=-\frac{1}{2\omega}\frac{\mathrm{d}^{2}}{\mathrm{d}x^{2}}+\frac{1}{\omega}V_{k}\left(x\right)=-\frac{1}{2\omega}\frac{\mathrm{d}^{2}}{\mathrm{d}x^{2}}+\frac{1}{2}\omega x^{2}-\frac{1}{\omega}\frac{\mathrm{d}^{2}}{\mathrm{d}x^{2}}\left[\ln{W\left(u_{1},\cdots ,u_{k}\right)}\right]	,	
	\end{equation}
which acquire the standard Schr\"{o}dinger form through the substitution
	\begin{equation}\label{47}
y=\sqrt{\omega}x	.
	\end{equation}

The $ \left(2k+1\right) $-th order ladder operators $ \mathcal{L}_{2k+1}^{\pm} $ and the Hamiltonian $ H_{k} $ (in fact a family) generate a $ 2k $-th order PHA. As we saw in section \ref{**Sec-3}, systems connected with the PIV equation must have third-order ladder operators and satisfy a second-order PHA. Thus, we need to identify the subfamily of $ H_{k} $ (if any) suitable for the purposes of this paper, i.e., having as well third-order differential ladder operators. In fact, there is a reduction theorem that will be useful later on, enabling us to reduce to second- the initial order ($ 2k $) of the PHA and thus leading to solutions of the PIV equation \cite{BF11a}.
\\

Let us analyze next some novelties arising in the complex case as compared with the real one. We will start by using the simplest situation, with $ k=1 $, for illustrating them. Let us note that the transformation function associated to the factorization energy $\varepsilon_1$ will be denoted as $u_1(x)$ throughout this Section, It will be taken either as one of the bound state eigenfunctions of Eqs.~(\ref{35}), (\ref{37}) or as the general solution of Eq.~(\ref{33}).

\subsection{First-order transformation}
As a first atypical fact, let us use as transformation function a bound state eigenfunction $ \phi_{i}\left(x\right) $ of the complex oscillator. Note that the parity of $ \phi_{i} $ is $ \left(-1\right)^{i} $, and the even eigenfunctions (with $ i $ even) do not have zeros on the real axis so they induce non-singular first-order SUSY transformations while the odd ones (with $ i $ odd) have one node at $ x=0 $, inducing transformations with a singularity at this point. Let us study next each of these two cases separately.

\subsubsection{Bound state even seed solution}\label{even}
\begin{figure}[!h]
	\centering
	\begin{tabular}{c c}
	\subfigure[]	{\includegraphics[width=0.45\textwidth]{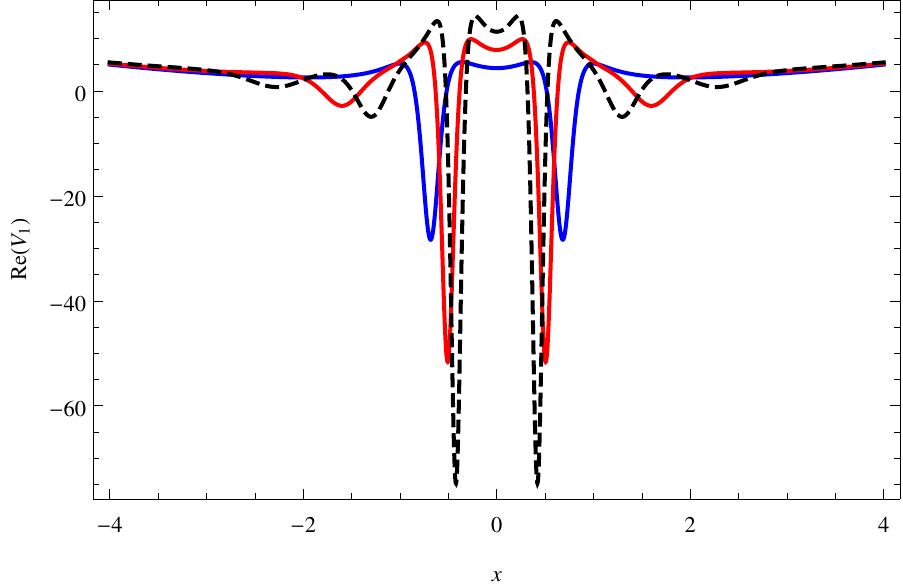}}	&
	\subfigure[]{\includegraphics*[width=0.45\textwidth]{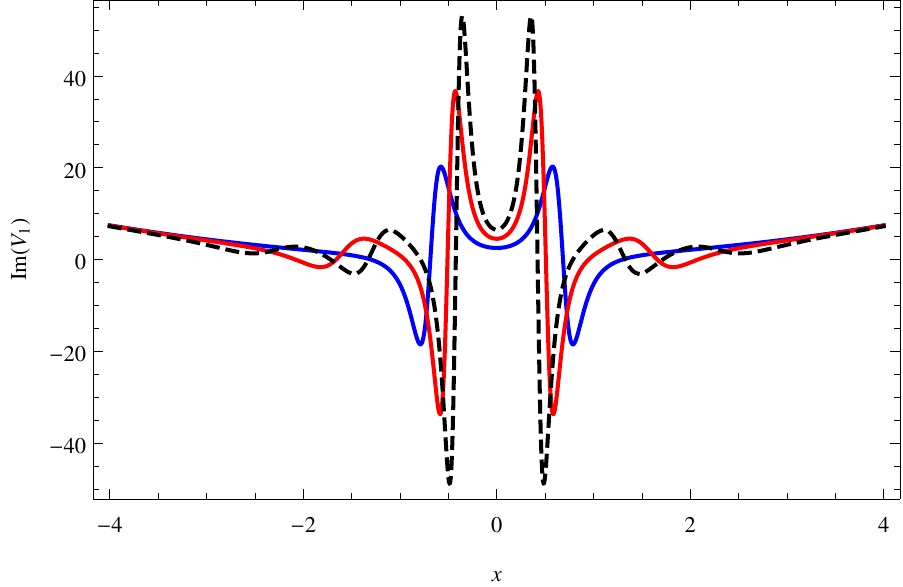}}	\\
	\end{tabular}
		\caption{Real (a) and imaginary parts (b) for the first-order SUSY partner potential $ V_{1} $ generated by an even eigenfunction of $ H_{0} $. Several non-trivial cases are shown for $ j=1 $ (blue), $ j=2 $ (red) and $ j=3 $ (dashed) with $ \theta =\frac{\pi}{6} $.}
		\label{--fig:3}
\end{figure}

Let us take as transformation function an even eigenfunction of $ H_{0} $:
	\begin{subequations}
	\begin{equation}\label{48a}
u_{1}\left(x\right)=\mathrm{H}_{2j}(\sqrt{\omega}x)e^{-\frac{1}{2}\omega x^{2}}	,	\quad	j=0,1,2,\cdots	.
	\end{equation}
The first-order SUSY partner potential of the complex oscillator turns out to be
	\begin{equation}\label{48b}
V_{1}\left(x\right)=\frac{1}{2}\omega^{2}x^{2}+\omega-8j\omega\left[\left(2j-1\right)\frac{\mathrm{H}_{2j-2}(\sqrt{\omega}x)}{\mathrm{H}_{2j}(\sqrt{\omega}x)}-2j\left(\frac{\mathrm{H}_{2j-1}(\sqrt{\omega}x)}{\mathrm{H}_{2j}(\sqrt{\omega}x)}\right)^{2}\right]	.
	\end{equation}
	\end{subequations}
In Fig. \ref{--fig:3} it is shown the typical behavior of $ V_{1} $ for several values of $ j $.

The explicit expressions for the eigenfunctions of $ H_{1} $ are given by
	\begin{equation}\label{49}
\psi_{n}^{\left(1\right)}\left(x\right)\propto\left[4j\frac{\mathrm{H}_{2j-1}(\sqrt{\omega}x)}{\mathrm{H}_{2j}(\sqrt{\omega}x)}\mathrm{H}_{n}(\sqrt{\omega}x)-2n\mathrm{H}_{n-1}(\sqrt{\omega}x)\right]e^{-\frac{1}{2}\omega x^{2}}	,	\;	n=0,1,2,\cdots	;	\;	n\neq 2j	,
	\end{equation}
so that the spectrum of $ H_{1} $ becomes:
	\begin{equation}\label{50}
\mathsf{Sp}(H_{1})=\left\lbrace\left(n+\frac{1}{2}\right)\omega ;\; n=0,1,2,\cdots ;\; n\neq 2j\right\rbrace	,
	\end{equation}
i.e., the eigenvalue $ E_{2j} $ of $ H_{0} $ has been deleted in order to generate $ H_{1} $, since the wavefunction annihilated by $ A_{1}^{-} $ in Eq.~(\ref{10}) is not square-integrable.
\\

As can be seen, for $ j>0 $ the spectrum of $ H_{1} $ is composed of two equally spaced energy ladders: a finite one of length $ 2j $, which starts from $ E_{0}=\frac{\omega}{2} $ and ends at $ E_{2j-1}=\left(2j-\frac{1}{2}\right)\omega $; an endless ladder departing from $ E_{2j+1}=\left(2j+\frac{3}{2}\right)\omega $. Let us note that the lower ends of both ladders supply us with two extremal energies as well as the associated extremal states for our system.
\\

\begin{figure}[!h]
	\centering
	\begin{tabular}{c c}
	\subfigure[]	{\includegraphics[width=0.45\textwidth]{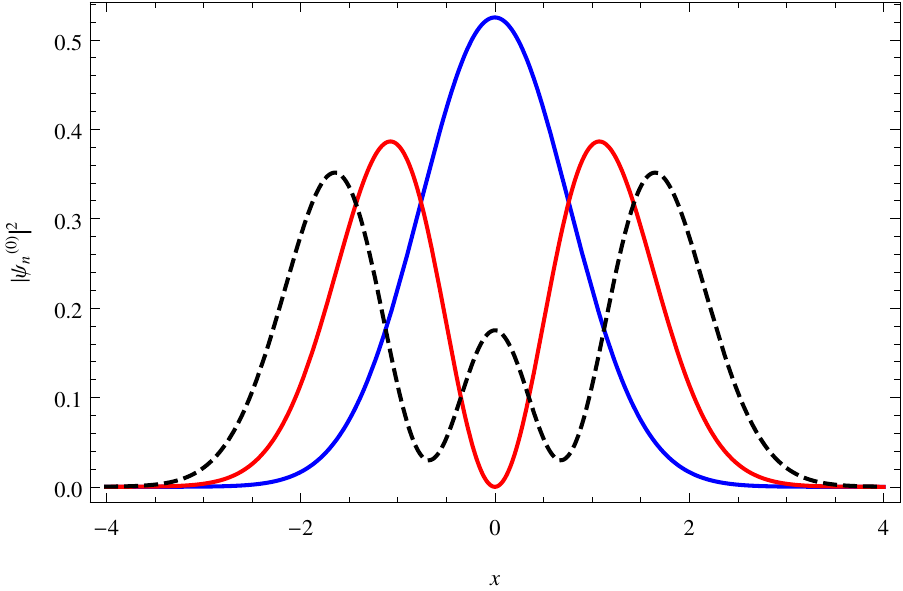}}	&
	\subfigure[]{\includegraphics*[width=0.45\textwidth]{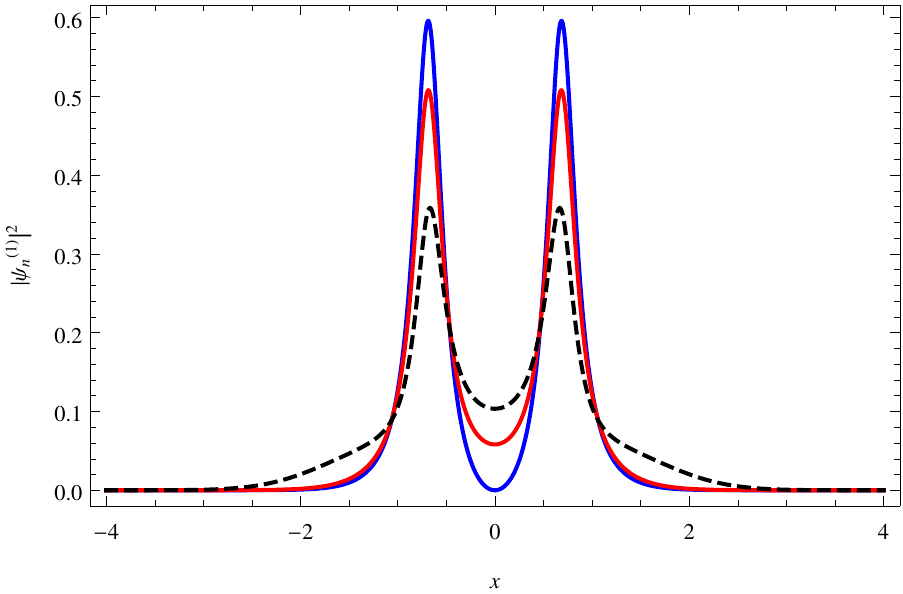}}	\\
	\end{tabular}
		\caption{Comparison between the normalized probability densities for: (a) the eigenstates of $ H_{0} $ with $ n=0 $ (blue), $ n=1 $ (red) and $ n=2 $ (dashed); (b) the transformed ones for $ n=0 $ (blue), $ n=1 $ (red) and $ n=3 $ (dashed). The transformation function is the first even excited state $ \phi_{2}\left(x\right) $ with $ \theta =\frac{\pi}{6} $.}
		\label{--fig:4}
\end{figure}

In Fig. \ref{--fig:4} it is compared the normalized probability densities for the eigenfunctions of $ H_{0} $ with the corresponding transformed ones for $ H_{1} $. The graphs correspond to the case when the transformation function is the first even excited state $ \phi_{2}\left(x\right) $ with $ \theta =\frac{\pi}{6} $.

\subsubsection{Bound state odd seed solution}
Suppose now that the transformation function is an odd eigenfunction of $ H_{0} $,
	\begin{subequations}
	\begin{equation}\label{51a}
u_{1}\left(x\right)=\mathrm{H}_{2j+1}(\sqrt{\omega}x)e^{-\frac{1}{2}\omega x^{2}}	,	\quad	j=0,1,2,\cdots	,
	\end{equation}
which vanishes at $ x=0 $ and thus induces a singularity at the same point.	Despite this fact, let us use the standard formulas of the first-order SUSY QM and analyze then the obtained results.
\\

The first-order SUSY partner potential reads now
	\begin{equation}\label{51b}
V_{1}\left(x\right)=\frac{1}{2}\omega^{2}x^{2}+\omega-4\left(2j+1\right)\omega\left[2j\frac{\mathrm{H}_{2j-1}(\sqrt{\omega}x)}{\mathrm{H}_{2j+1}(\sqrt{\omega}x)}-\left(2j+1\right)\left(\frac{\mathrm{H}_{2j}(\sqrt{\omega}x)}{\mathrm{H}_{2j+1}(\sqrt{\omega}x)}\right)^{2}\right]	,
	\end{equation}
while the possible eigenfunctions of $ H_{1} $ are given by
	\begin{equation}\label{51c}
\psi_{n}^{\left(1\right)}\left(x\right)\propto\left[\left(4j+2\right)\frac{\mathrm{H}_{2j}(\sqrt{\omega}x)}{\mathrm{H}_{2j+1}(\sqrt{\omega}x)}\mathrm{H}_{n}(\sqrt{\omega}x)-2n\mathrm{H}_{n-1}(\sqrt{\omega}x)\right]e^{-\frac{1}{2}\omega x^{2}}	,	\;	n=0,1,2,\cdots	;	\;	n\neq 2j+1	.
	\end{equation}
	\end{subequations}
The potential $ V_{1}\left(x\right) $ has a singularity at $ x=0 $, which can be interpreted as coming from a complex oscillator potential with an infinite barrier at the origin. Hence, the coordinate domain has to be redefined between two equivalent options: either $ \left(-\infty ,0\right) $ or $ \left(0,\infty\right) $. For convenience, from now on we will take the domain $ x\in\left(0,\infty\right) $.	\\
The complex oscillator potential with an infinite barrier at the origin has odd parity eigenfunctions, which vanish at $ x=0 $ and $ x=\infty $. On the other hand, the transformed eigenfunctions (\ref{51c}) do not always satisfy the same conditions \cite{FM14}, thus we need to select the appropriate subset. In fact, it turns out that the $ \psi_{n}^{\left(1\right)} $ with $ n $ odd vanishes at $ x=0 $ and $ x=\infty $ while for $ n $ even it diverges at $ x=0 $. Thus, the eigenfunctions of $ H_{1} $ in the restricted domain $ \left(0,\infty\right) $ are indeed:
	\begin{subequations}
	\begin{equation}\label{52a}	
\psi_{n}^{\left(1\right)}\left(x\right)\propto\left[\left(4j+2\right)\frac{\mathrm{H}_{2j}(\sqrt{\omega}x)}{\mathrm{H}_{2j+1}(\sqrt{\omega}x)}\mathrm{H}_{2n+1}(\sqrt{\omega}x)-\left(4n+2\right)\mathrm{H}_{2n}(\sqrt{\omega}x)\right]e^{-\frac{1}{2}\omega x^{2}}	,	\;	n=0,1,2,\cdots	;	\;	n\neq j	.
	\end{equation}
The spectrum of $ H_{1} $ becomes now:
	\begin{equation}\label{52b}
\mathsf{Sp}(H_{1})=\left\lbrace E_{n}=\left(2n+\frac{3}{2}\right)\omega ;\; n=0,1,2,\cdots ;\; n\neq j\right\rbrace	,
	\end{equation}
	\end{subequations}
i.e., the eigenvalue $ \left(2j+\frac{3}{2}\right)\omega $ of the complex oscillator with an infinite barrier at the origin has been deleted for generating the potential $ V_{1}\left(x\right) $, which domain is now $ x\in\left(0,\infty\right) $.
\\

Similarly as for the non-singular case of section \ref{even}, the spectrum of $ H_{1} $ can be seen as composed of two equally spaced energy ladders, but the spacing is now twice the original one for the non-singular case. In addition, the finite ladder has just $ j $ steps, starting from $ E_{0}=\frac{3}{2}\omega $ and ending at $ E_{j-1}=\left(2j-\frac{1}{2}\right)\omega $. On the other hand, the infinite ladder starts from $ E_{j+1}=\left(2j+\frac{7}{2}\right)\omega $.
\\

\begin{figure}[!h]
	\centering
	\begin{tabular}{c c}
	\subfigure[]	{\includegraphics[width=0.45\textwidth]{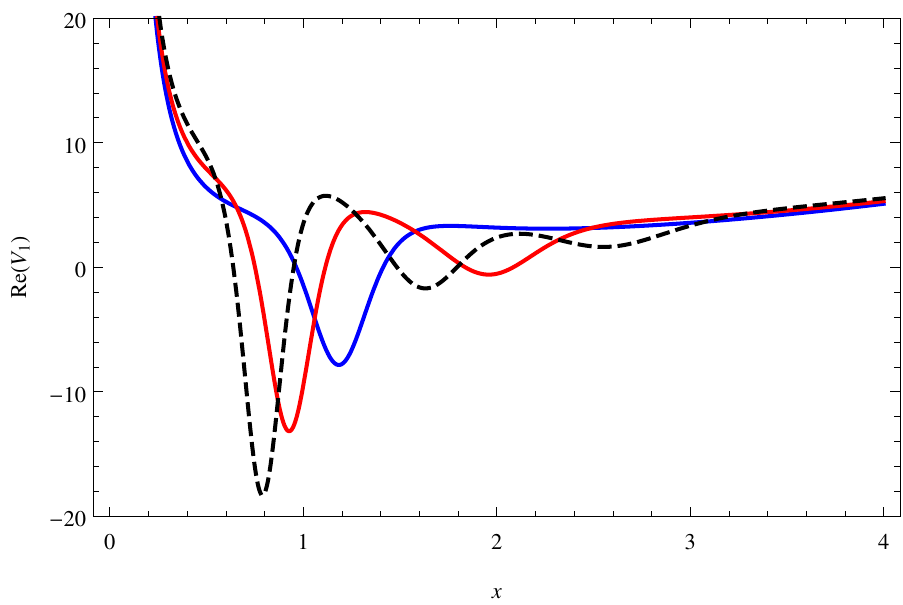}}	&
	\subfigure[]{\includegraphics*[width=0.45\textwidth]{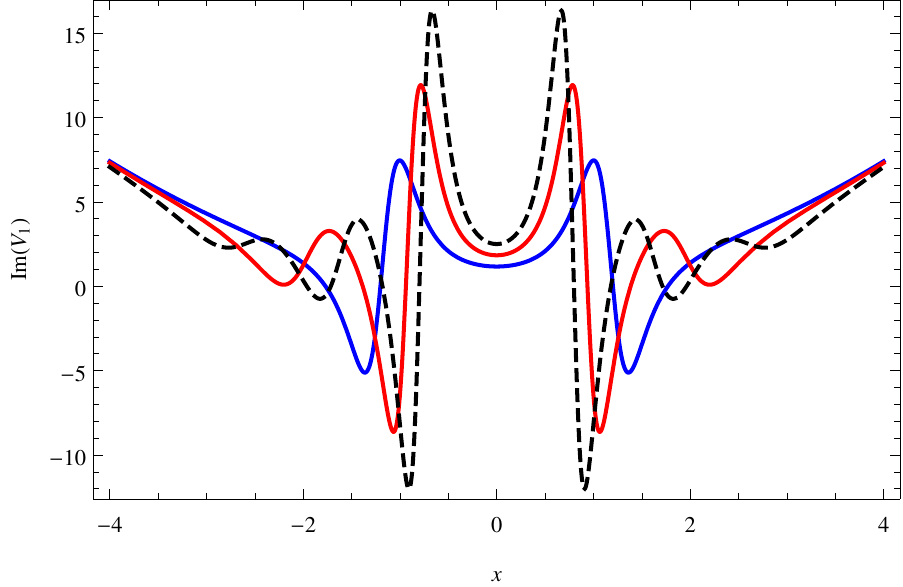}}	\\
	\end{tabular}
		\caption{Real (a) and imaginary parts (b) for the singular first-order SUSY partner potential $ V_{1} $ generated by an odd eigenfunction of $ H_{0} $. Several cases are shown for $ j=1 $ (blue), $ j=2 $ (red) and $ j=3 $ (dashed) with $ \theta =\frac{\pi}{6} $.}
		\label{--fig:5}
\end{figure}

\begin{figure}[!h]
	\centering
	\begin{tabular}{c c}
	\subfigure[]	{\includegraphics[width=0.45\textwidth]{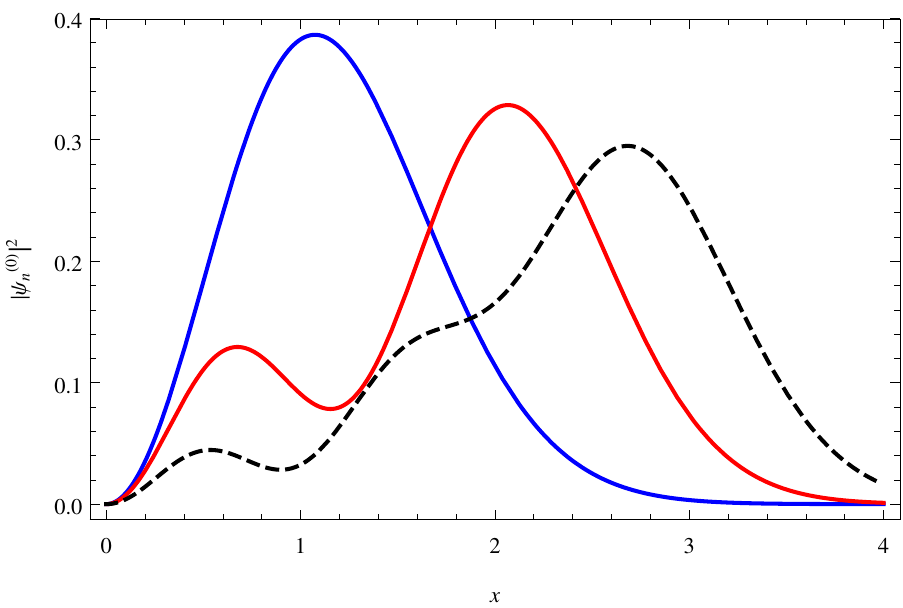}}	&
	\subfigure[]{\includegraphics*[width=0.45\textwidth]{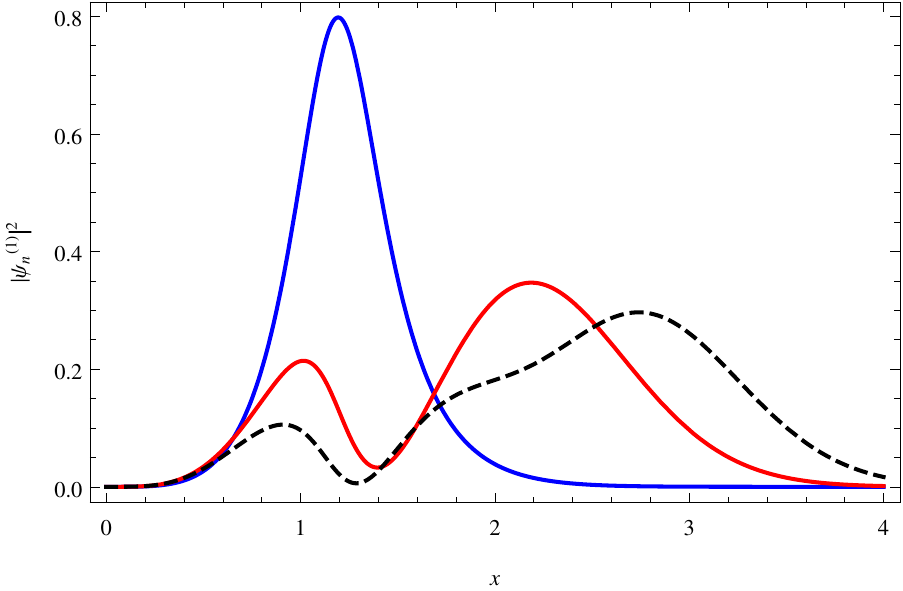}}	\\
	\end{tabular}
		\caption{Comparison between the normalized probability densities for: (a) the eigenstates of $ H_{0} $ with $ n=0 $ (blue), $ n=1 $ (red) and $ n=2 $ (dashed); (b) the transformed ones for $ n=0 $ (blue), $ n=2 $ (red) and $ n=3 $ (dashed). The transformation function is the odd excited state $ \phi_{3}\left(x\right) $ with $ \theta =\frac{\pi}{6} $.}
		\label{--fig:6}
\end{figure}

In Fig. \ref{--fig:5} the typical singular behavior of $ V_{1} $ for several values of $ j $ is shown while in Fig. \ref{--fig:6} the normalized probability densities for the eigenfunctions of $ H_{0} $ and $ H_{1} $ in the restricted domain $ x\in\left(0,\infty\right) $ are compared.

\subsubsection{General seed solution}
\begin{figure}[!h]
	\centering
	\begin{tabular}{c c}
	\subfigure[]	{\includegraphics[width=0.45\textwidth]{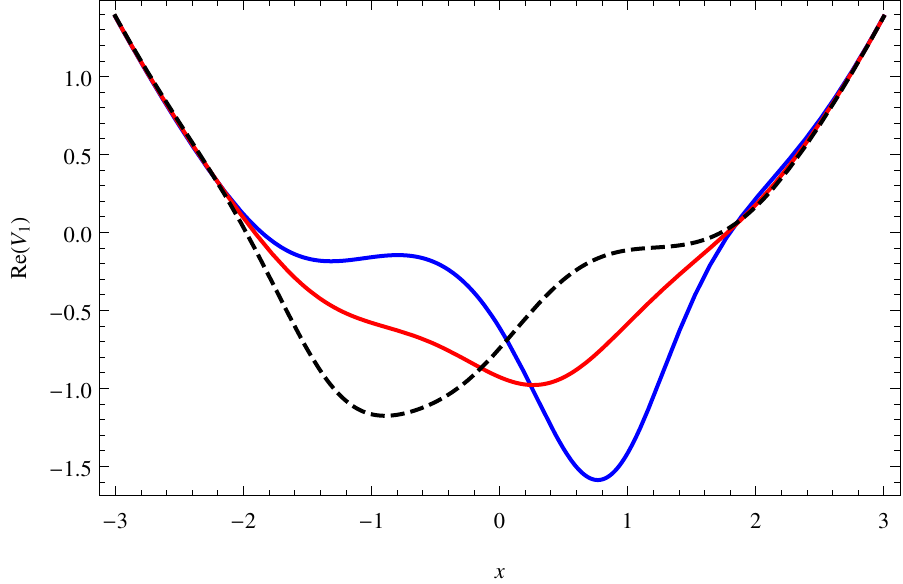}}	&
	\subfigure[]{\includegraphics*[width=0.45\textwidth]{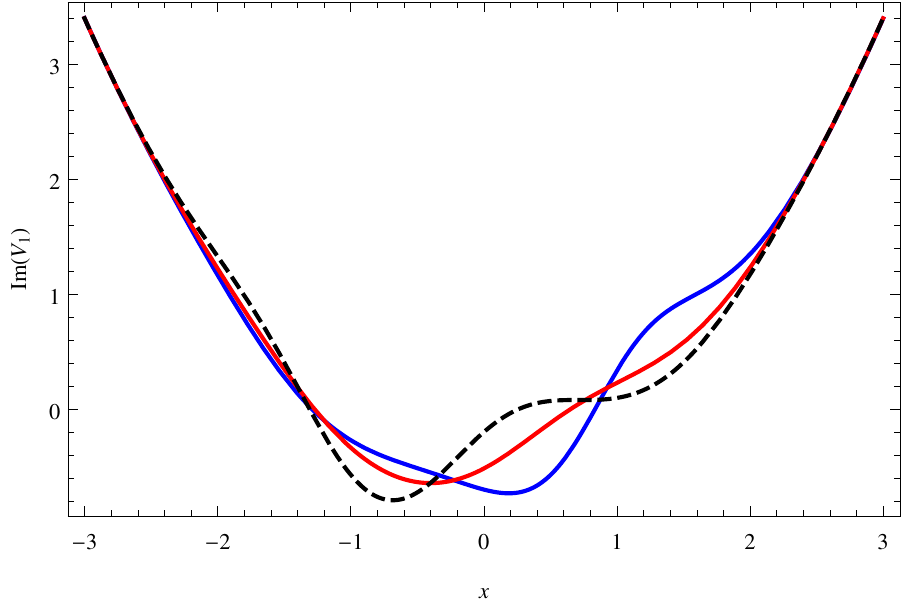}}	\\
	\end{tabular}
		\caption{Real (a) and imaginary parts (b) for the potential $ V_{1} $ of Eq.~(\ref{55}). Several cases are shown for $ \nu_1 =-0.6+0.3i $ (blue), $ \nu_1 =0.3i $ (red) and $ \nu_1 =0.6+0.3i $ (dashed). We have taken $ \varepsilon_1 =-\frac{\omega}{2} $ and $ \theta =\frac{\pi}{6} $.}
		\label{--fig:7}
\end{figure}

\begin{figure}[!h]
	\centering
	\begin{tabular}{c c}
	\subfigure[]	{\includegraphics[width=0.45\textwidth]{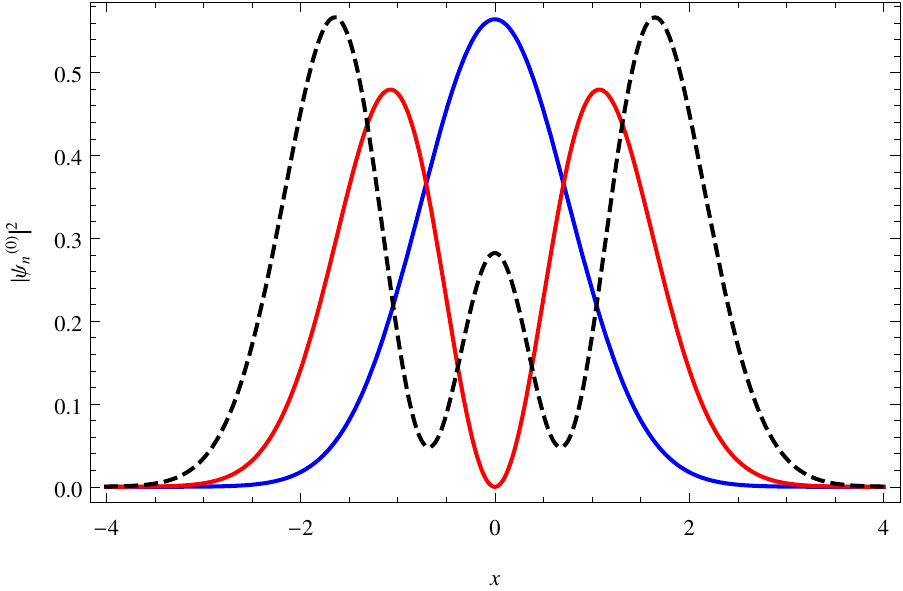}}	&
	\subfigure[]{\includegraphics*[width=0.45\textwidth]{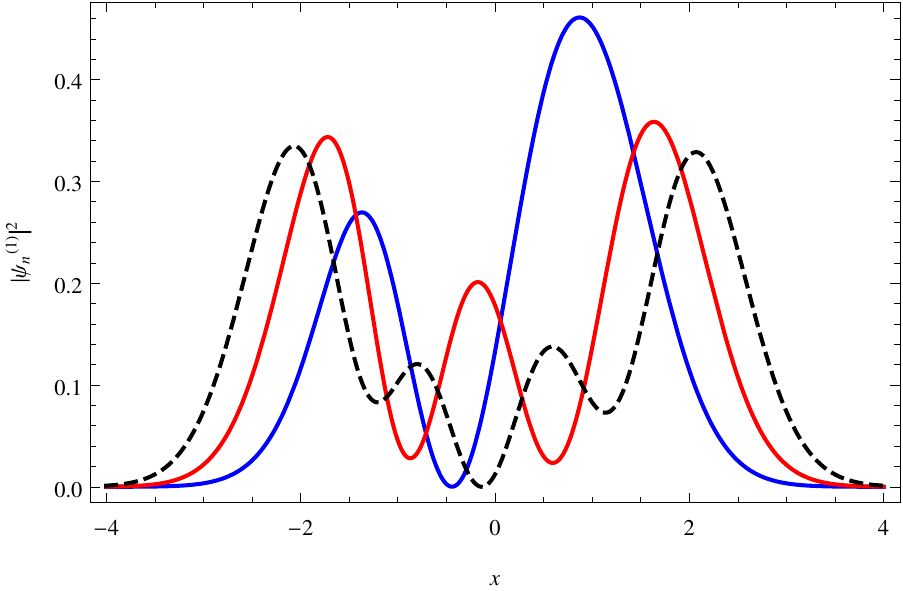}}	\\
	\end{tabular}
		\caption{Comparison between the normalized probability densities for: (a) the eigenstates of $ H_{0} $ with $ n=0 $ (blue), $ n=1 $ (red) and $ n=2 $ (dashed) and (b) their corresponding quantities for $ H_{1} $. We have taken $ \varepsilon_1 =-\frac{\omega}{2} $, $ \theta =\frac{\pi}{6} $ and $ \nu_1 =0.6+0.3i $.}
		\label{--fig:8}
\end{figure}

For a transformation function $ u_{1}\left(x\right)\equiv u\left(x,\varepsilon_{1}\right) $ of the form given in Eq.~(\ref{33}) with $ \varepsilon_{1}\in\mathbb{C} $, the first-order SUSY partner potential of the complex oscillator turns out to be:
	\begin{equation}\label{53}
V_{1}\left(x\right)=\frac{1}{2}\omega^{2}x^{2}-\frac{u_{1}^{\prime\prime}\left(x\right)}{u_{1}\left(x\right)}+\left[\frac{u_{1}^{\prime}\left(x\right)}{u_{1}\left(x\right)}\right]^{2}	.
	\end{equation}
For some particular values of $ \varepsilon_1$ there is a further simplification since the confluent hypergeometric function reduces to a polynomial. The simplest example appears for $ \varepsilon_1 =-\frac{\omega}{2} $ for which
	\begin{equation}\label{54}
u_1\left(x\right)=e^{\frac{1}{2}\omega x^{2}}\left[1+\frac{\nu_1}{\sqrt{\omega}}\mathtt{erf}(\sqrt{\omega}x)\right]	.
	\end{equation}
In this case the new potentials become:
	\begin{equation}\label{55}
V_{1}\left(x\right)=\frac{1}{2}\omega^{2}x^{2}-\omega -\frac{\mathrm{d}}{\mathrm{d}x}\left[\frac{2\nu_1}{\sqrt{\pi}}\frac{e^{-\omega x^{2}}}{1+\frac{\nu_1}{\sqrt{\omega}}\mathtt{erf}(\sqrt{\omega}x)}\right]	,
	\end{equation}
which are complex extensions of the Abraham-Moses-Mielnik potentials derived several years ago \cite{AM81,Mie84}. The corresponding spectrum is composed of the levels $ E_{n}=\left(n+\frac{1}{2}\right)\omega $, but now $ n=-1,0,1,\cdots $. A plot of the real and imaginary parts of $ V_{1} $ for several values of $ \nu_1$ are shown in Fig. \ref{--fig:7}. The associated normalized probability densities for some eigenstates with a fixed value of $ \nu_1$ are plotted in Fig. \ref{--fig:8}.

\subsection{Higher-order SUSY partners}
\begin{figure}[!h]
	\centering
	\begin{tabular}{cc}
	\subfigure[]	{\includegraphics[width=0.42\textwidth]{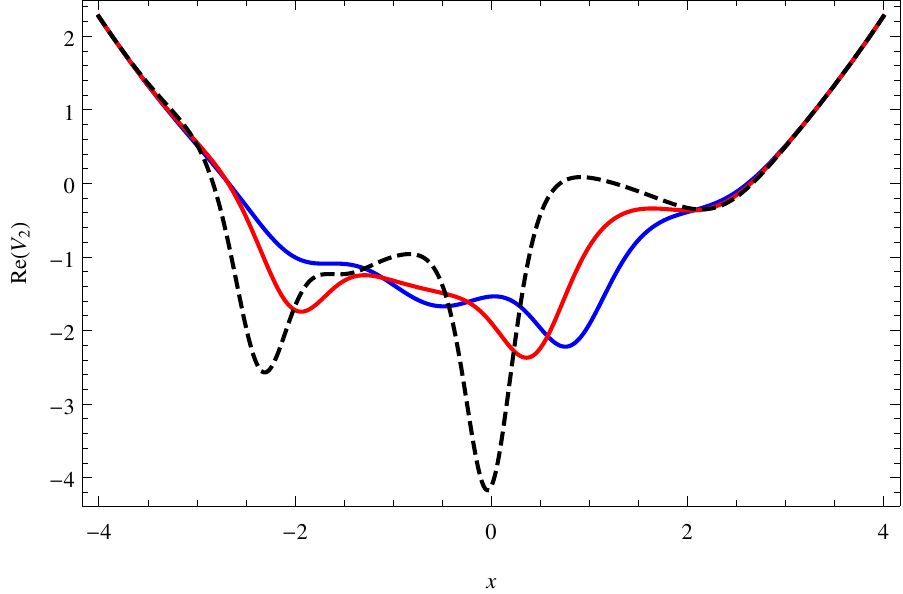}}	&
	\subfigure[]{\includegraphics[width=0.42\textwidth]{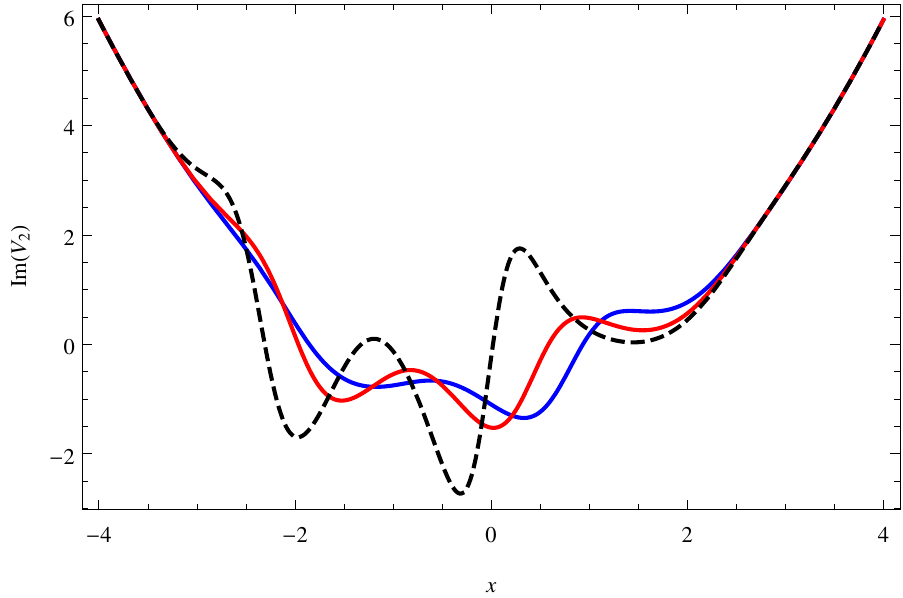}}	\\
	\end{tabular}
		\caption{SUSY partners $ V_{2} $ of Eq.~(\ref{57}). Real (a) and imaginary parts (b) for several values of the parameter $ \nu_{1} $: $ \nu_{1}=0.1+0.4i $ (blue), $ \nu_{1}=0.5+0.4i $ (red) and $ \nu_{1}=0.9+0.4i $ (dashed). We have taken $ \varepsilon_{1}=-\frac{\omega}{2} $ and $ \theta =\frac{\pi}{6} $.}
		\label{--fig:9}
\end{figure}

\begin{figure}[!h]
	\centering
	\begin{tabular}{cc}
	\subfigure[]	{\includegraphics[width=0.42\textwidth]{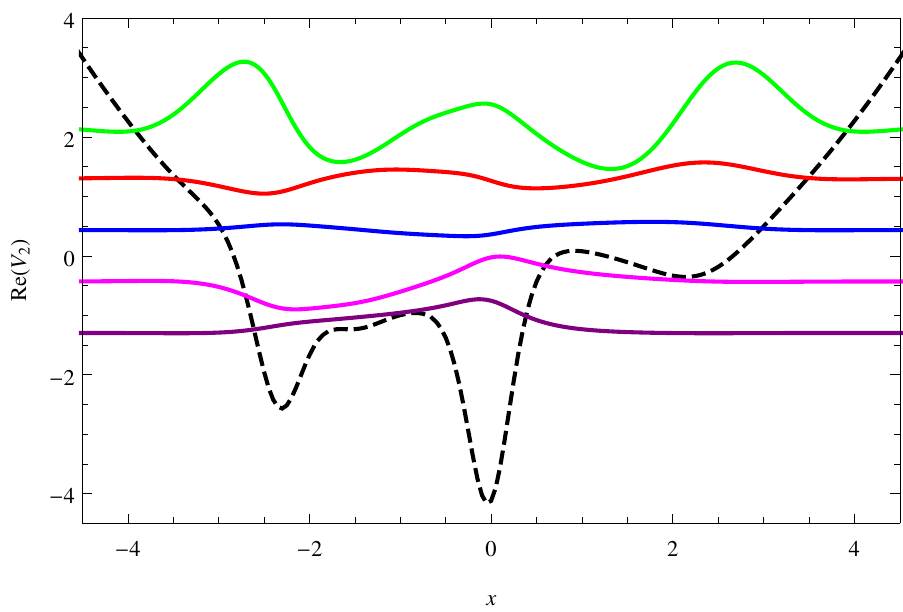}}	&
	\subfigure[]	{\includegraphics*[width=0.42\textwidth]{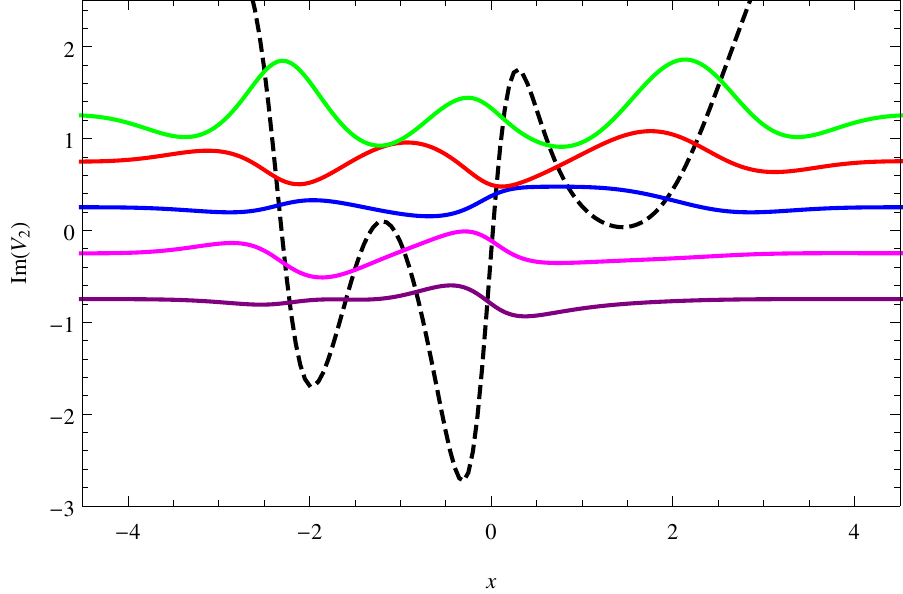}}	\\
	\end{tabular}
		\caption{(a) Real part of $ V_{2} $ (dashed) together with the real part of $ \psi_{n}^{\left(2\right)} $ for $ n=0 $ (blue), $ n=1 $ (red) and $ n=2 $ (green). The magenta and purple lines represent the real parts of the new eigenstates $ \psi_{-1}^{\left(2\right)}\;(=\psi_{\varepsilon_{1}}^{\left(2\right)}) $ and $ \psi_{-2}^{\left(2\right)}\;(=\psi_{\varepsilon_{2}}^{\left(2\right)}) $ respectively. (b) The same as in (a) but for the imaginary part of $ V_{2} $ and the corresponding eigenfunctions. We have taken $ \varepsilon_{1}=-\frac{\omega}{2} $, $ \theta =\frac{\pi}{6} $ and $ \nu_{1}=0.9+0.4i $.}
		\label{--fig:10}
\end{figure}

A $ k $-th order SUSY transformation requires to choose in general, $ k $ independent seed solutions of the Schr\"{o}dinger equation.	In order to induce the reduction theorem for linking with the PIV equation, however, we will consider just connected seed solutions satisfying:
	\begin{subequations}
	\begin{align}
u_{j}\left(x\right)	&	=	\left(a^{-}_{\omega}\right)^{j-1}u_{1}\left(x\right)	,	\label{56a}	\\
\varepsilon_{j}	&	=	\varepsilon_{1}-\left(j-1\right)\omega	,	\quad	j=1,\cdots ,k,	\label{56b}
	\end{align}
	\end{subequations}
with $ u_{1}(x) $ being a complex seed solution associated to $ \varepsilon_{1} $. As in the first order case, $u_{1}(x)$ can be taken either as one of the eigenfunctions of Eqs.~(\ref{35}), (\ref{37}), as the general Schr\"odinger solution of Eq.~(\ref{33}), or as the solution given in Eq.~(\ref{54}). In particular, for $ k=2 $ the second-order SUSY partner potential of the complex oscillator, for a fixed factorization energy $ \varepsilon_{1} $, acquires the form:
	\begin{equation}\label{57}
V_{2}\left(x\right)=\frac{1}{2}\omega^{2}x^{2}-\frac{W^{\prime\prime}}{W}+\left(\frac{W^{\prime}}{W}\right)^{2}	,
	\end{equation}
where
	\begin{equation}
W=:W(u_{1},a_{\omega}^{-}u_{1})\propto\left(\omega^{2}x^{2}+\omega -2\varepsilon_{1}\right)u_{1}^{2}-(u_{1}^{\prime})^{2}	.	\label{58}
	\end{equation}
In Fig. \ref{--fig:9} we have plotted the real and imaginary parts of $ V_{2}$, for the seed solution of Eq.~(\ref{54}) with $ \varepsilon_{1}=-\frac{\omega}{2} $ and several values of $ \nu_{1} $. Some eigenfunctions of $ H_{2} $, $ \lbrace\psi_{n}^{\left(2\right)},\; n=-2,-1,0,\cdots\rbrace $, for $ \varepsilon_{1}=-\frac{\omega}{2} $ and $ \nu_{1}=0.9+0.4i $, are shown in Fig. \ref{--fig:10}.

\section{Connection with the Painlev\'{e} IV equation}\label{**Sec-5}
We have seen previously that a second-order PHA with third-order ladder operators describes systems which are linked with the PIV equation (see section \ref{**Sec-3}). Thus, a first-order SUSY transformation applied to the complex oscillator provides solutions to the PIV equation. In fact, from Eq.~(\ref{30c}) we can obtain solutions to such equation in terms of the extremal states of the system by making carefully the changes from the harmonic to the complex oscillator:
	\begin{equation}\label{59}
g_{1}^{\left(j\right)}\left(x\right)=-\sqrt{\omega}x-\frac{1}{\sqrt{\omega}}\left[\ln{\psi_{\mathcal{E}_{j}}\left(x\right)}\right]^{\prime}	,	\quad	j=1,2,3,
	\end{equation}
where the subscript is used to indicate the order of the SUSY transformation. The labeling of the wavefunctions in Eqs. (30) is not essential, so that by making cyclic permutations of the indices we can generate three different solutions to the PIV equation. Do not forget that $ u_{1} $ is a solution (not necessarily physical) of the Schr\"{o}dinger equation (\ref{7}) associated to the factorization energy $ \varepsilon_{1} $. Thus, for fixed value of $ \varepsilon_{1} $ the solutions to the PIV equation constitute in fact a 3-parametric family. The three extremal states (up to constant factors) and their corresponding energies are given by
	\begin{subequations}
	\begin{alignat}{3}
\psi_{\mathcal{E}_{1}}	&	\propto	A_{1}^{+}\psi_{0}^{\left(0\right)}\propto A_{1}^{+}e^{-\frac{1}{2}\omega x^{2}}	,	&&	\qquad\qquad	\mathcal{E}_{1}	&&	=	\frac{\omega}{2}	,	\label{60a}	\\
\psi_{\mathcal{E}_{2}}	&	\propto	\frac{1}{u_{1}}	,	&&	\qquad\qquad	\mathcal{E}_{2}	&&	=	\varepsilon_{1}	,	\label{60b}	\\
\psi_{\mathcal{E}_{3}}	&	\propto	A_{1}^{+}a_{\omega}^{+}u_{1}	,	&&	\qquad\qquad	\mathcal{E}_{3}	&&	=	\varepsilon_{1}+\omega	.	\label{60c}
	\end{alignat}
	\end{subequations}

\begin{figure}[!ht]
	\centering
	\begin{tabular}{c c}
	\subfigure[]	{\includegraphics[width=0.42\textwidth]{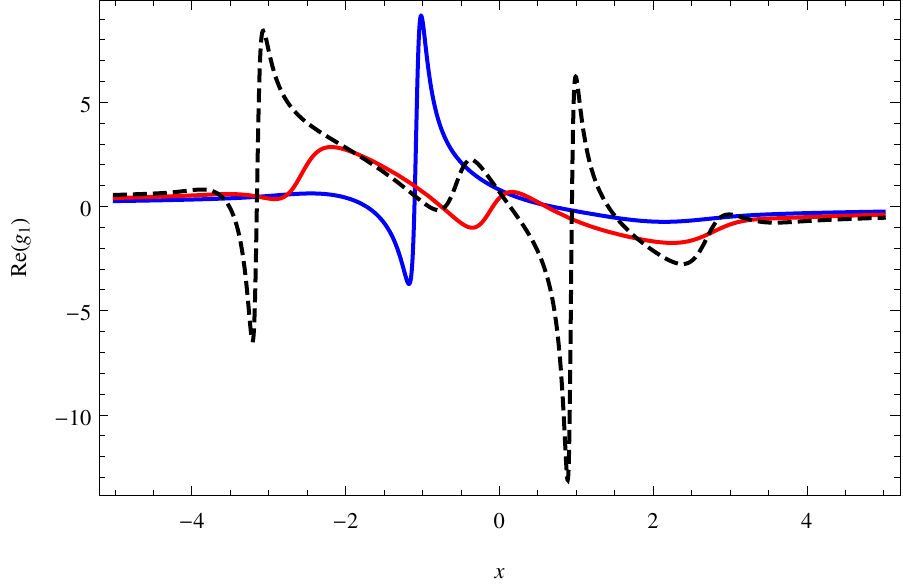}}
	&	\subfigure[]	{\includegraphics[width=0.42\textwidth]{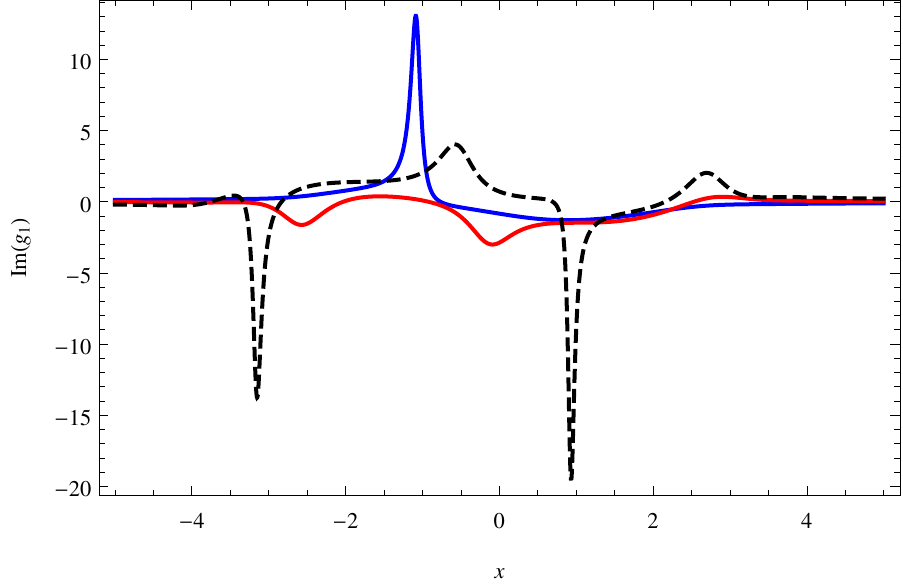}}	\\
	\end{tabular}
		\caption{Solution $ g_{1}^{\left(2\right)} $ to the PIV equation. Real (a) and imaginary parts (b) of $ g_{1}^{\left(2\right)} $ for $ \varepsilon_{1}=0.01+i $ (blue), $ \varepsilon_{1}=1+i $ (red) and $ \varepsilon_{1}=2+i $ (dashed). We have taken $ \theta =\frac{\pi}{6} $ and $ \nu =0.8+0.5i $.}
		\label{--fig:11}
\end{figure}

In Fig. \ref{--fig:11} we show a set of complex solutions to the PIV equation for several values of the factorization energy and a fixed $ \nu $.

\begin{figure}[!ht]
	\centering
	\includegraphics[width=0.4\textwidth]{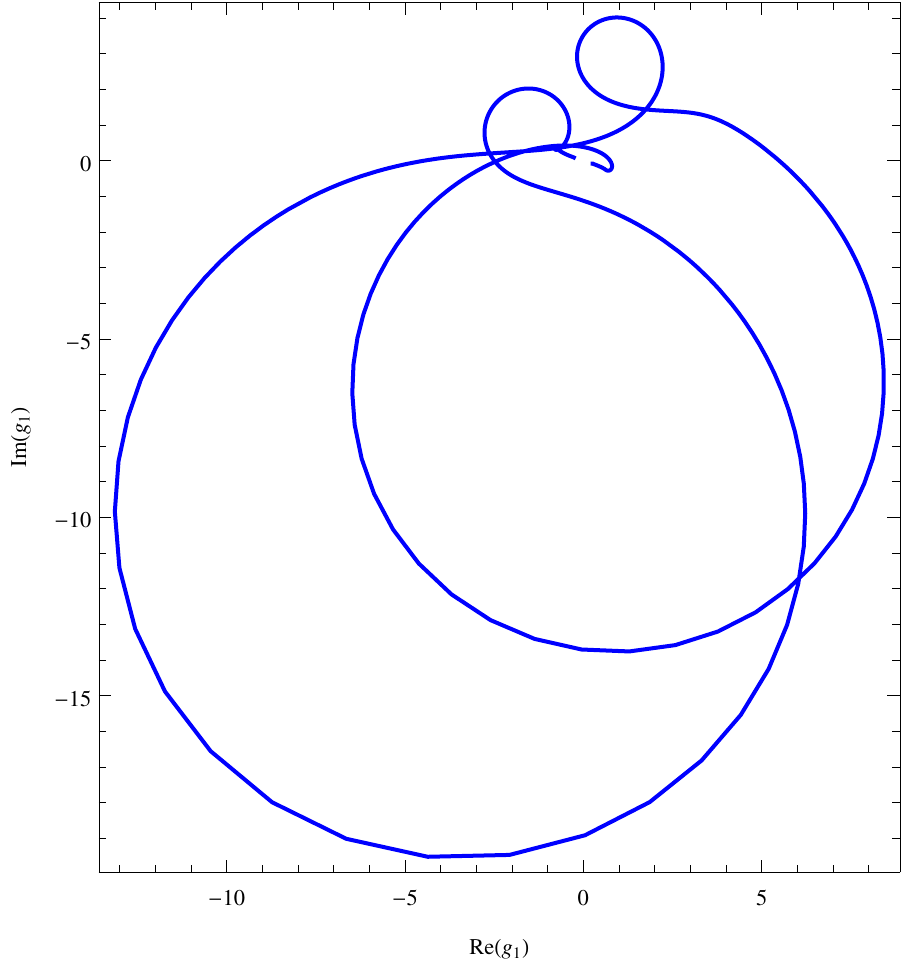}
		\caption{Parametric plot of $ \mathrm{Re}{(g_{1}^{\left(2\right)})} $ versus $ \mathrm{Im}{(g_{1}^{\left(2\right)})} $. The function $ g_{1}^{\left(2\right)} $ tends to $ \left(0,0\right) $ in the limits $ \left| x\right|\rightarrow\infty $. We have taken $ \varepsilon =2+i $, $ \theta =\frac{\pi}{6} $, $ \nu =0.8+0.5i $ and $ \left| x\right| <10 $.}
		\label{--fig:12}
\end{figure}

If $ \psi_{\mathcal{E}_{j}}\neq 0 $ $ \forall\;x\in\mathbb{R} $, the functions $ g_{1}^{\left(j\right)} $ are non-singular. Furthermore, they have a null asymptotic behavior, $ g_{1}^{\left(j\right)}\rightarrow 0 $ when $ \vert x\vert\rightarrow\infty $. This can be seen from a different viewpoint in Fig. \ref{--fig:12}: for large values ​​of $ \vert x\vert $, the curve tends slowly to close around $ \mathrm{Re}{(g_{1}^{\left(2\right)})}=\mathrm{Im}{(g_{1}^{\left(2\right)})}=0 $.

\begin{figure}[!ht]
	\centering
	\begin{tabular}{c c}
	\subfigure[]{\includegraphics[width=0.42\textwidth]{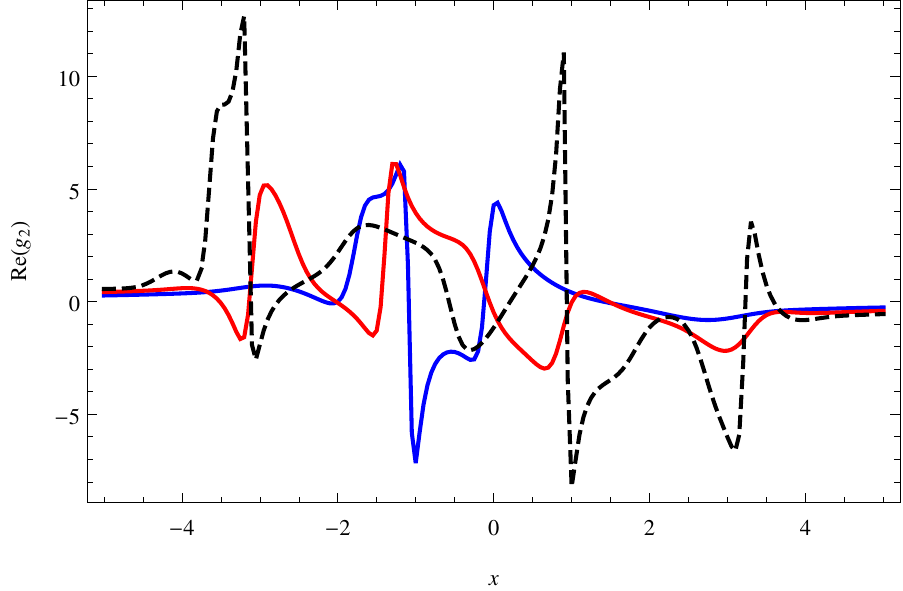}}
	&	\subfigure[]{\includegraphics[width=0.42\textwidth]{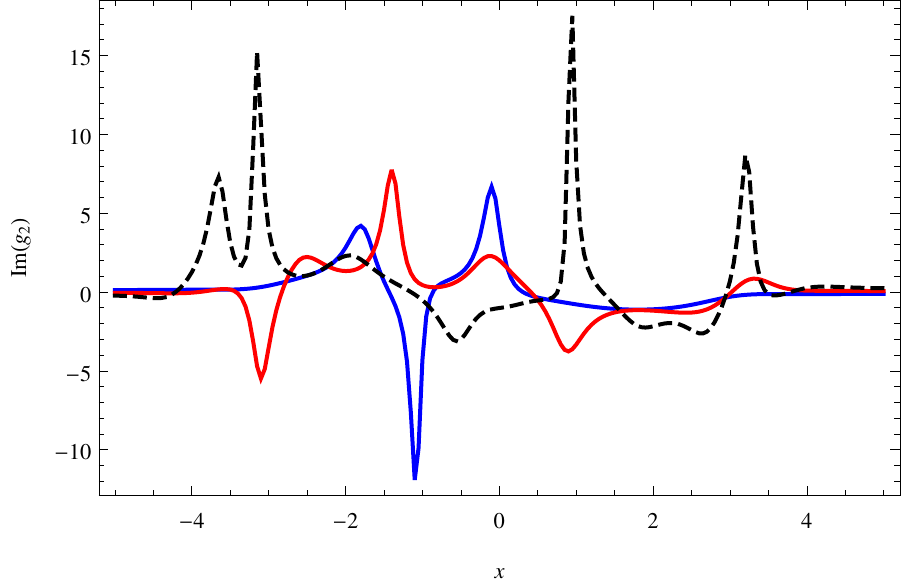}}	\\
	\end{tabular}
		\caption{Solution $ g_{2} $ to the PIV equation. Real (a) and imaginary (b) parts of $ g_{2} $ for $ \varepsilon_{1}=0.01+i $ (blue), $ \varepsilon_{1}=1+i $ (red) and $ \varepsilon_{1}=2+i $ (dashed). We have taken $ \theta =\frac{\pi}{6} $ and $ \nu_{1} =0.8+0.5i $.}
		\label{--fig:13}
\end{figure}

For higher orders of SUSY $ \left(k>1\right) $, the solution to the PIV equation appears after the reduction theorem is produced. In fact, one of these solutions is given by
	\begin{equation}\label{61}
g_{k}\left(x\right)=-\sqrt{\omega}x-\frac{1}{\sqrt{\omega}}\left[\ln{\frac{W\left(u_{1},\cdots ,u_{k-1}\right)}{W\left(u_{1},\cdots ,u_{k}\right)}}\right]^{\prime}	,	\quad	k\geq 2	,
	\end{equation}
where the $ u_{j} $, $ j=1,\cdots ,k $ satisfy Eqs.~(56). In Fig. \ref{--fig:13} we illustrate the typical behavior of $ g_{2} $ for several values of $ \varepsilon_{1} $ and $ \nu_{1}=0.8+0.5i $.
\\

It is worth to note that in the harmonic oscillator limit,  $ \mathrm{Im}{\left(\varepsilon_{1}\right)}=\mathrm{Im}{\left(\nu_{1}\right)}=\theta =0 $, it is possible to recover the results reported in \cite{BF11a}, where the number of nodes of the non-singular $ g $ functions for $ \varepsilon_{1}<\frac{1}{2} $ are fixed by the order of the SUSY transformation, and they tend to zero when $ \vert x\vert\rightarrow\infty $.

\section{Conclusions}\label{**Sec-6}
In this paper we have introduced a complex generalization of the standard factorization method to analyze the SUSY partners of the non-Hermitian Hamiltonian associated to the complex oscillator. The corresponding potentials were generated through SUSY transformations involving complex factorization energies. Some interesting transformations, which cannot be implemented in the real case without introducing undesired singularities in the new potentials, were found. In particular, the first-order SUSY transformation that uses bound state eigenfunctions can be implemented to obtain either non-singular potentials whose energy levels become the initial ones (but for the employed bound state energy) or potentials with a singularity at $ x=0 $, which are almost isospectral to the complex oscillator with an infinite barrier at the origin. In addition, it was identified a complex extension for the Abraham-Moses-Mielnik potentials, which turn out to be isospectral to the initial complex oscillator (up to an energy displacement).
\\

Let us note that the physical interpretation of complex energies remains open, although possible applications related to absorbent (dissipative) systems have been noticed recently \cite{FGR07a,FR08} (see also \cite{Si73}).
\\

In order to finish, several additional open problems can be pointed out. In the first place, it is not clear the way we have to move on the complex energy plane in order to obtain a natural classification of the complex SUSY transformations, the associated potentials and the corresponding solutions to the PIV equation. In addition, this time we did not obtain any heuristic criterion to link the number of zeros of the PIV solution with the order of the SUSY transformation. We are convinced that these two subjects, although difficult, are worth to be explored in the future for clarifying better the peculiarities of the complex SUSY transformations discussed in this paper.

\section*{Acknowledgments}
The authors acknowledge the support of Conacyt Mexico, project 152574. JCG also acknowledges the
Conacyt PhD scholarship 262669.


\end{document}